\documentclass[11pt]{article}

\pdfoutput=1 
\usepackage{amssymb}
\usepackage{amsmath}
\usepackage{bbold}
\usepackage{amstext}
\usepackage{graphicx,epsfig}
\usepackage{epsfig}
\usepackage{verbatim} 
\usepackage{fancybox}
\usepackage{xcolor}
\usepackage{color}
\usepackage{ulem}
\usepackage{enumitem}
\usepackage{subfigure}
\usepackage{bbm}
\usepackage{parskip}
\usepackage{dsfont}
\usepackage[numbers,sort&compress]{natbib}
\usepackage{ytableau}
\usepackage{framed}
\usepackage{hyperref}
\usepackage[title]{appendix}
\usepackage[]{mdframed}

\newcommand{\rd}{{\rm d}}

\newcommand{\Mp}{M_{\rm Pl}}

\definecolor{Green}{rgb}{0, 0.65, 0.31}
\newcommand{\SW}[1]{\textcolor{Green}{[SW: #1]}}

\author{Toshifumi Noumi, Sam S. C. Wong}
\date{\today}

\begin{document}

\rightline{UT-Komaba/26-3}

\bigskip

\def\thefootnote{\fnsymbol{footnote}}

\begin{center}
\LARGE{\textbf{Extremal Love: tidal/electromagnetic deformability, logarithmic running and the weak gravity conjecture}} \\[0.5cm]
 
\large{Toshifumi Noumi$^{1}$\footnote{\href{mailto:}{\texttt{tnoumi@g.ecc.u-tokyo.ac.jp}}} and Sam S. C. Wong,$^{2}$\footnote{\href{mailto:}{\texttt{samwong@cityu.edu.hk}}}}
\\[0.5cm]

\small{
\textit{
$^1$Graduate School of Arts and Sciences, University of Tokyo, \\
Komaba, Meguro-ku, Tokyo 153-8902,
Japan\\
 \vskip 5pt
~~$^2$Department of Physics, City University of Hong Kong, \\
Tat Chee Avenue, Kowloon, Hong Kong SAR, China}
 }

\vspace{.2cm}

\end{center}

\vspace{.6cm}

 \vspace{0.2cm}
\centerline{\small{\bf Abstract}}
%\vspace{-0.2cm}
{\small\noindent 
In General Relativity, the tidal Love numbers of black holes vanish, implying they are resistant to tidal deformation. This ``rigidity" is easily broken in the presence of higher-derivative corrections. Focusing on extremal charged black holes in Einstein-Maxwell EFT, we compute the static linear response for both the vector ($\ell=1$) and parity-odd tensor ($\ell \ge 2$) sectors. We find that the resulting tidal Love numbers are non-zero and exhibit logarithmic running, a hallmark of quantum corrections. Crucially, we show that the sign of these deformations is not arbitrary; the induced electric and magnetic susceptibilities and their log runnings in the $\ell=1$ sector are constrained by unitarity and the Weak Gravity Conjecture. Furthermore, due to gravito-electromagnetic mixing, we find the cross log runnings and show that they are the same, which we explain through the worldline effective field theory.

}

\vspace{0.3cm}
\noindent
\def\thefootnote{\arabic{footnote}}
\setcounter{footnote}{0}

\section{Introduction}

Tidal interactions provide a powerful probe into the internal structure and response of compact objects. In theories of gravity, the tidal deformability of an object is quantified by its tidal Love numbers (TLNs)~\cite{poisson_will_2014}, originally introduced in the context of Newtonian theory and later extended to general relativity (GR). These parameters, which measure the induced tidal field in response to external fields, have attracted considerable attention because of their potential observability in gravitational wave signals from compact binary coalescence \cite{Flanagan:2007ix,ET:2019dnz,Reitze:2019iox,Kalogera:2021bya,Branchesi:2023mws}. 

Surprisingly, for all black holes described by GR, including Einstein-Maxwell theory, the static linear Love numbers are known to vanish in four dimensions~\cite{Binnington:2009bb,Damour:2009vw,Damour:2009va,Kol:2011vg,Pani:2015hfa,Pani:2015nua,Gurlebeck:2015xpa,Porto:2016zng,LeTiec:2020spy, Chia:2020yla,LeTiec:2020bos}, leading to the intriguing notion that black holes are ``rigid” or ``non-deformable" in some sense. This ``non-deformability" is a reflection of the simplicity and universality of black hole solutions, analogous to the no-hair theorems. It has been suggested that this apparent ``naturalness problem" in GR~\cite{Porto:2016zng} can be connected to symmetries associated with perturbations around black holes in GR~\cite{Hui:2020xxx,Charalambous:2021kcz,Charalambous:2021mea,Hui:2022vbh,Charalambous:2022rre,Ivanov:2022qqt,Katagiri:2022vyz}. It has also been suggested that it has an origin in conformal field theories~\cite{Bonelli:2021uvf,Kehagias:2022ndy, Lupsasca:2025pnt}.

However, this property is broken in higher dimensions~\cite{Kol:2011vg,Cardoso:2019vof,Hui:2020xxx,Pereniguez:2021xcj,Charalambous:2023jgq,Rodriguez:2023xjd,Hadad:2024lsf,Xia:2025zfp}, by environmental effects surrounding black holes~\cite{Baumann:2018vus,Cardoso:2019upw,DeLuca:2021ite,DeLuca:2022xlz,Katagiri:2023yzm}, in scenarios of exotic black hole mimickers~\cite{Pani:2015tga, Cardoso:2017cfl,Herdeiro:2020kba,Chen:2023vet,Berti:2024moe}, or within the framework of modified gravity~\cite{Cardoso:2017cfl,Cardoso:2018ptl,Cvetic:2021vxa,DeLuca:2022tkm}.
Furthermore, in our pursuit of a fundamental theory of the universe, GR is not the ultimate answer. At high energies, phenomena such as new particles or tiny extra dimensions may emerge, hinting at a richer landscape of physics beyond GR and the standard model of particles. In the effective field theory (EFT) approach to gravity, these effects manifest as higher derivative operators. Such terms systematically capture a broad range of corrections, including quantum effects and contributions from physics at high energy scales. When these corrections are taken into account, a richer phenomenology emerges. 

In this work, we study the tidal deformations of extremal (electric) black holes within the Einstein-Maxwell EFT, augmented by the leading order four derivative operators. Extremal black holes are of special interest as they lie at the boundary of cosmic censorship~\cite{Horowitz:2016ezu,Shiu:2016weq,Crisford:2017gsb,Yu:2018eqq} and often provide probes of quantum gravity effects, such as the Weak Gravity Conjecture (WGC)~\cite{Arkani-Hamed:2006emk}. Higher derivative corrections to these black holes are bounded by the WGC~\cite{Kats:2006xp,Cheung:2018cwt,Hamada:2018dde} and they can induce non-trivial tidal responses. Several studies have investigated deformed extremal black holes, focusing on specific operators, particular parity sectors, or individual harmonic modes $\ell$~\cite{DeLuca:2022tkm,DeLuca:2022xlz,Barbosa:2025uau,DiRusso:2025qpf}. Our analysis begins with the dipolar ($\ell=1$) sector, a regime that has received limited attention despite its direct connection to electromagnetic susceptibility. We then extend our study to the parity-odd $\ell \ge 2$ sector, providing explicit results for the $\ell=2$ and $\ell=3$ modes.

We compute the corrections to the vector (electromagnetic) and tensor (gravitational) tidal responses due to higher derivative terms—parameterized by Wilson coefficients $\alpha_i$. Crucially, these macroscopic deformations are not arbitrary; we show that they are intimately linked to fundamental constraints on the UV theory. Specifically, we demonstrate that the positivity bounds imposed on the EFT coefficients by unitarity and the Weak Gravity Conjecture translate into specific constraints on the black hole's electric and magnetic susceptibilities.

Additionally, our bulk calculations uncover the presence of logarithmic running in the tidal response. This behavior, often associated with the renormalization group flow in the point-particle effective theory, suggests that the ``Love numbers" in this context are scale-dependent quantities~\cite{Goldberger:2004jt,Kol:2007rx,Kol:2007bc,Goldberger:2009qd,Ivanov:2022hlo,Barbosa:2025uau,Garcia-Saenz:2025urd}. We discuss how this running is consistent with the expectations of effective field theory and how unitarity dictates the direction of this flow.

A central feature of our study is the treatment of gravito-electromagnetic mixing. For charged objects, gravitational and electromagnetic responses cannot be separated; a pure gravitational tidal field can induce an electromagnetic multipole, and conversely, an external magnetic field can generate a gravitational response. We perform a detailed perturbative analysis of these cross-responses for $\ell \ge 2$ perturbations. Interestingly, we confirm that this mixing is symmetric in the gauge invariant logarithmic corrections: the gravitationally induced magnetic log running is identical to the magnetically induced gravitational log running. We further elucidate this structure by considering the worldline EFT, showing that it arises naturally from the limited set of allowed interaction operators on the worldline.

The paper is organized as follows. In Section II, we review the Einstein-Maxwell EFT setup, outlining the four-derivative operators and the corrected geometry of extremal black holes. Section III discusses electromagnetic susceptibilities in standard Maxwell theory and details their matching to a worldline EFT. The core analysis of tidal deformations for the corrected extremal black hole is presented in Sections IV and V. In Section IV, we analyze the $\ell=1$ electromagnetic response and derive constraints imposed by the WGC and unitarity. Furthermore, in Section V, we investigate the $\ell \ge 2$ sector, incorporating both magnetic and parity-odd gravitational perturbations; here, we uncover the gravito-electromagnetic mixing and the emergence of logarithmic corrections in the tidal response. Finally, we conclude in Section VI with a discussion.

\section{Einstein-Maxwell EFT up to four derivative operators}

\subsection*{EFT setup}

The theory of interest is\footnote{ 
 The complete set of of our derivative operators is given in \cite{Kats:2006xp}. The chosen set of operators is the minimal set such that the complete set can be induced by a field redefinition \cite{Cheung:2014ega,Cheung:2018cwt}. Furthermore, since the scattering amplitude is field redefinition invariant, unitary constraints are imposed on the chosen set of operators~\cite{Cheung:2014ega,Hamada:2018dde,Bellazzini:2019xts}. } 
\begin{align} \label{eqn:action}
S=\int d^4x\sqrt{-g}
\bigg[
\frac{1}{2 \kappa^2}R-\frac{1}{4}F_{\mu\nu}F^{\mu\nu}
+ \alpha_1 \kappa^4(F_{\mu\nu}F^{\mu\nu})^2
+ \alpha_2 \kappa^4 (F_{\mu\nu}\widetilde{F}^{\mu\nu})^2
+ \alpha_3 \kappa^2  F_{\mu\nu}F_{\rho\sigma}W^{\mu\nu\rho\sigma}
\bigg]
\end{align}
where $\kappa= \Mp^{-1}$ and $W^{\mu\nu\rho\sigma}$ is the Weyl tensor,
\begin{align}
  W_{\mu\nu\rho\sigma} = R_{\mu\nu\rho\sigma} + \frac{2}{d-2} \left( R_{\mu [\sigma }g_{\rho]\nu} - R_{\nu [\rho }g_{\sigma]\mu}\right) + \frac{2}{(d-1)(d-2)}R g_{\mu [\rho }g_{\sigma] \nu} \,,
\end{align}
where $d$ is the number of dimensions. 

A version of the weak gravity conjecture implies that the mass of extremal black holes is increased by higher derivative corrections. Requiring this for extremal dyonic black holes with the electromagnetic charges $(Q_e,Q_m)=(Q\cos\varphi ,Q\sin\varphi)$ leads to a one-parameter family of positivity bounds on the effective couplings~$\alpha_i$~\cite{Cheung:2019cwi}:
\begin{align}
\alpha_1\cos^2 2\varphi+\alpha_2\sin^2 2\varphi\geq\frac{\alpha_3}{4}\cos 2\varphi\,.
\end{align}
In particular, the bounds for $\varphi=0,\frac{\pi}{4},\frac{\pi}{2}$ require
\begin{align}~\label{eqn:WGCconstraints}
\alpha_1\geq \frac{1}{4}|\alpha_3|\,,
\quad
\alpha_2\geq0\,.
\end{align}
Suggestively, the same bounds are obtained from consistency of scattering amplitudes as long as Planck-suppressed quantum gravitational corrections are negligible (see, e.g.,~\cite{Hamada:2018dde,Bellazzini:2019xts}).\footnote{
Recently, the connection between this version of the weak gravity conjecture and causality was extended in~\cite{Abe:2025vdj} to all orders of the derivative expansion within the context of nonlinear electrodynamics. It would be interesting to extend our analysis to nonlinear electrodynamics to further explore causality constraints on Love.
}

%The weak gravity conjecture requires that 
%\begin{align}
% 4\alpha_1 -\alpha_3 >0,
%\end{align}
%\cite{Kats:2006xp} while positivity of scattering amplitude in $4D$ \cite{Hamada:2018dde} requires that 
%\begin{align}
% \alpha_1>0,\quad \alpha_2>0,
%\end{align} 
%and  positivity of scattering amplitude in $3D$ (after a compactification along $z$) \cite{Bellazzini:2019xts}  requires that 
%\begin{align}
% 4 \alpha_{1} > |\alpha_3|, \quad \alpha_2>0
%\end{align}

\subsection*{Extremal black holes}

The corrected extremal black hole due to the four derivative operators in \eqref{eqn:action} is given by 
\begin{align} \label{eqn:BHmetric}
    ds^2 =- f_t(r) dt^2 + \frac{1}{f_r(r)}dr^2 + r^2 d\Omega^2, \quad F = \frac{1}{2}F_{tr}(r) dt \wedge dr,
\end{align}
where $f_t(r)$, $f_r(r)$ and the electric field $F_{tr}$ are corrected as 
\begin{align}
    &f_t(r) \nonumber\\
    & = \left( 1- \frac{r_h}{r}\right)^2 \left[ 1+\frac{4 \kappa ^2 (7 \alpha_3-48 \alpha_1)}{15  r_h  r }+\frac{16 \kappa ^2 (\alpha_3-9 \alpha_1)}{15 r^2}+\frac{4 \kappa ^2 r_h (\alpha_3-24 \alpha_1)}{15 r^3}-\frac{4 \kappa ^2 r_h^2 (12 \alpha_1+7 \alpha_3)}{15 r^4}\right]  \nonumber \\
    &f_r(r)  \nonumber\\
    &=  \left( 1- \frac{r_h}{r}\right)^2 \left[1+\frac{4 \kappa ^2 (7 \alpha_3-48 \alpha_1)}{15  r_h r }+\frac{16 \kappa ^2 (\alpha_3-9 \alpha_1)}{15 r^2}+\frac{4 \kappa ^2 r_h (\alpha_3-24 \alpha_1)}{15 r^3}-\frac{16 \kappa ^2 r_h^2 (\alpha_1+\alpha_3)}{5 r^4}\right]  \nonumber \\
    &F_{tr} = \frac{\sqrt{2} r_h}{\kappa r^2} \left( 1+ \frac{4 \kappa ^2 (6 \alpha_1-\alpha_3)}{3 r_h^2}+\frac{16 \alpha_3 \kappa ^2 r_h}{r^3}-\frac{2 \left(\kappa ^2 r_h^2 (48 \alpha_1+23 \alpha_3)\right)}{3 r^4}\right) 
\end{align}
We have chosen the integration constants in the first-order solutions so that the horizon radius $r_h$ remains fixed, which simplifies the imposition of boundary conditions for small tidal perturbations at the horizon. As a result, our solution differs slightly from that in~\cite{Kats:2006xp}; however, the two are related by an ${\cal O}(\alpha)$ redefinition of $r_h$, under which the charge-to-mass ratio remains invariant.
\begin{align}
     |Q| &= \frac{1}{4\pi} \int \star F = \frac{\sqrt{2} r_h}{\kappa} \left(1 + \frac{4  \kappa ^2 (6 \alpha_1-\alpha_3)}{3 r_h^2} \right), \\
     M &= r_h \left( 1 + \frac{2   \kappa ^2 (48 \alpha_1-7 \alpha_3)}{15r_h^2} \right).
\end{align}
Therefore the corrected charge to mass ratio is given by, 
\begin{align}
  \frac{|Q|}{M} = \frac{\sqrt{2}}{\kappa}\left( 1 + \frac{2 \kappa ^2}{5r_h^2} ( 4\alpha_1 - \alpha_3) \right),
\end{align}
which agrees with the standard result.

\section{Vector Love numbers and susceptibility} 
In this section we review the basic properties about linear materials in electromagnetism and define the corresponding vector Love numbers. Let's start with a massless vector field in a spherically symmetric spacetime, which is described by the usual Maxwell theory, $    S = \int \rd^4x \sqrt{|g|} \left(  -\frac{1}{4} F_{\mu\nu}F^{\mu\nu} \right). $ 
Utilizing the spherical symmetry of the background, the gauge field $A_{\mu}$ can be decomposed as 
\begin{align} \label{eqn:Adecomp}
    A_\mu = \sum_{\ell, m} \left(\begin{array}{c}
a_{0, \ell m}  \\
a_{r, \ell m} \\
a^{(L)}_{\ell m} \nabla_A   +a^{(T)}_{\ell m} \epsilon_A^{\; B} \nabla_B  
\end{array}\right) Y^m_{\ell}
\end{align}
where $a_{0, \ell m}, a_{r, \ell m}, a^{(L)}_{\ell m}$ and $a^{(T)}_{\ell m}$ are functions of $(t,r)$ and we will suppress the $\ell, m $ indices from now on, $\nabla_A$ and $\epsilon_A^{\; B}$ are the covariant derivative and the Levi-Civita symbol on the two sphere respectively. Under parity transform $(\theta, \phi) \to (\pi -\theta, \phi+\pi)$ on the two sphere, the scalar and vector spherical harmonics pick up the factor
\begin{align}
  Y^m_{\ell}, \;  \nabla_A Y^m_{\ell} &: (-)^{\ell}, \nonumber \\
  \epsilon_A^{\; B} \nabla_B Y^m_{\ell} & : (-)^{\ell+1}.
\end{align}
Therefore, the parity even sector $(a_0, a_r, a^{(L)})$ decouple from the parity odd sector $a^{(T)}$ given that the theory is parity invariant.

\subsection{Point particle effective action} \label{sec:3.1}
In order to compare it with a flat space object, one usually considers the point particle effective action,
\begin{align} \label{eqn:ppeft}
    S = \int \rd^4x \sqrt{|g|} \left(  -\frac{1}{4} F_{\mu\nu}F^{\mu\nu} \right) + &\int \rd \tau e \bigg[ \frac{1}{2 e^2} \dot{x}^{\mu}\dot{x}_{\mu} -\frac{m^2}{2}+\frac{q}{e} \dot{x}^{\mu}A_{\mu} \nonumber \\
    &+ \sum_{\ell=1} \frac{1}{2 \ell!}\left( \lambda^{(E)}_{\ell} E_{\langle \ell \rangle}E^{\langle \ell \rangle} + \frac{\lambda^{(B)}_{\ell}}{2}B_{\langle \ell \rangle}B^{\langle \ell \rangle} \right) \bigg],
\end{align}
with the symmetric trace free tensors (denoted as $\langle \dots\rangle  $)  $E_{\langle \ell \rangle}$ and $B_{\langle \ell \rangle}$ defined as
\begin{align}
  E_{\langle \ell \rangle}  = \partial_{ \langle a_1}\dots \partial_{a_{\ell-1}} E_{a_{\ell} \rangle }, \quad B_{\langle \ell \rangle}  = \partial_{ \langle a_1}\dots \partial_{a_{\ell-1}} B_{a_{\ell} \rangle }.
\end{align}
\footnote{We only include the parity conserving worldline operators. If parity is broken, one can include the following operator as well 
\begin{align}
    \sum_{\ell=1} \frac{1}{ \ell!} \lambda^{(E B)}_{\ell}E_{\langle \ell \rangle}B^{\langle \ell \rangle}  \,. 
\end{align}}
In the above action, $m$ and $q$ are the mass and charge of the point particle, and $\lambda^{(E)}$, $\lambda^{(B)}$ are the electromagnetic (vector) Love numbers that characterize the linear electric polarizability and the magnetic susceptibility \cite{Hui:2020xxx}. The indices $a_i$  are all properly contracted such that the action is a scalar under change of coordinate.

The equation of motion for this worldline EFT is 
\begin{align} \label{eqn:EFTeom}
    \partial_{\mu} F^{\mu \nu} = J^{\nu}_{\rm free} -\sum_{\ell}\lambda^{E}_{(\ell)} \frac{(-1)^\ell}{\ell!}  \partial_{\mu} & \int  d\tau\, \big[  \partial_{a_1}\dots \partial_{a_{\ell-1}}\big( \delta^{(4)}(x-z(\tau))  E^{\langle \ell-1 \rangle [ \mu} \big) u^{\nu]}   \nonumber \\
    & \qquad + \partial_{a_1}\dots \partial_{a_{\ell-1}}\big( \delta^{(4)}(x-z(\tau))  B^{\langle \ell-1 \rangle  \rho} \epsilon^{\mu\nu}_{\quad \rho\sigma} \big) u^{\sigma}  \big ],
\end{align}
where $z(\tau)$ is the worldline of the particle and $u^{\mu}$ is its four velocity, $J_{\rm free}^{\nu}$ denotes the free current describing the point charge. 
 
\subsection*{Electric susceptibility}
In the electric sector of the above theory, if the zeroth order boundary condition at spatial infinity is a static external electric field given by (in spatial cartesian coordinates)
\begin{align}
    A_0^{(0)} = c_{a_1\dots a_\ell} x^{a_1}\cdots x^{a_\ell}, 
\end{align}
the static electric potential at leading order in $\lambda$, $A_0 = A_0^{(0)}+ \lambda^{(E)}_{\ell} A_0^{(1)} $ can be solved (in the usual gauge choice $\dot{A_i}=0$, and in spherical coordinates) as 
\begin{align} \label{eqn:faltA0}
    A_0(\vec{x}) =  c^{E}_{\ell m}  Y^m_{\ell}(\theta, \phi)  \left[r^\ell + \lambda^{(E)}_{\ell} \frac{(2\ell -1)!! }{ 4\pi } r^{-\ell -1}  \right],
\end{align}
where the growing term $r^{\ell}$ with coefficient $ c^{E}_{\ell}$ quantifies the $\ell$-th multipole external electric field and the second term $r^{-\ell-1}$ with coefficient $c^{E}_{\ell}\lambda^{(E)}_{\ell}$ is the induced multipole due to the linear response of the point-particle. 

\begin{figure}[h]
    \centering
    \includegraphics[width=0.17\linewidth]{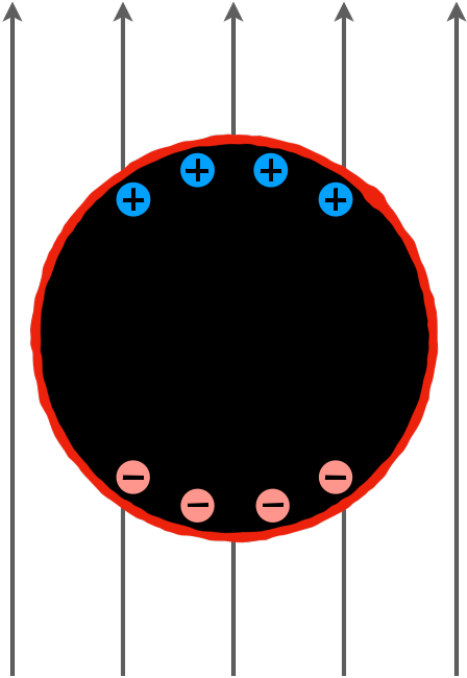}
    \caption{Positive electric polarizability aligns with the intuition that an external electric field displaces positive charges in its direction, as shown in the figure. This redistribution of charge  screens the electric field within the medium.}
    \label{fig:E_polar}
\end{figure}

Importantly, the sign of $ \lambda^{(E)}_{\ell}$ is not arbitrary. One can do an EFT matching with a sphere of linear dielectric material, with 
\begin{align}
    \vec{P} = \chi_E \vec{E}.
\end{align} 
The electric potential for a sphere, of radius $R$, of this material placed in any external electric field is obtained by  solving  $\nabla \cdot \vec{D} = \nabla \cdot( \epsilon_0 \vec{E}+ \vec{P} ) =0  $ ,
\begin{align} \label{eqn:EMmultipole}
    A_{0,\,{\rm outside}} &= \sum_{\ell, \, m} V_{\ell m} \left( r^\ell - \frac{\chi_E \ell R^\ell}{2 \ell +1 + \ell \chi_E  } \left(\frac{R}{r}\right)^{\ell +1} \right) Y_{\ell m}( \theta, \varphi) \,; \nonumber \\
     A_{0,\,{\rm inside}} &= \sum_{\ell, \, m} V_{\ell m} \frac{2\ell +1}{2\ell +1 + \ell \chi_E }r^\ell Y_{\ell m}( \theta, \varphi) \,.
\end{align}
In an effective description we match $ A_{0,\,{\rm outside} } $ with \eqref{eqn:faltA0} such that 
\begin{align}
     \lambda^{(E)}_{\ell} = -\frac{ 4\pi \ell   }{(2\ell -1)!!(2 \ell +1 + \ell \chi_E)} \chi_E R^{2\ell +1}.
\end{align}
Intuitively we always have {\it positive susceptibility} \footnote{We are aware that there exist materials with effective negative susceptibility. However, they are not isolated systems and external energy input is required to maintain negative susceptibility.}, $\chi>0$, for isolated systems therefore one expects that 
\begin{align}
     \lambda^{(E)}_{\ell} <0
\end{align} 
for ``healthy" deformation, as depicted in figure~\ref{fig:E_polar}. The microscopic picture is intuitive since electric dipoles in the material will align with the electric field to screen off the electric field inside the material. A useful gauge invariant quantity is the radial electric field $E_r=F_{0r}$
\begin{align} \label{eqn:radialE}
    E_r = -\sum_{\ell, \, m} V_{\ell m} \left(\ell r^{\ell-1} + \frac{\chi_E \ell(\ell+1) R^{\ell-1}}{2 \ell +1 + \ell \chi_E  } \left(\frac{R}{r}\right)^{\ell +2} \right) Y_{\ell m}( \theta, \varphi) ,
\end{align}
where we will compare the induced dipole electric field in later section. 

\subsection*{Magnetic susceptibility} 
Similarly, we should compare the effective theory with linear diamagnetic materials, with the magnetization $\vec{M}$ being related to $\vec{B}$ linearly through
\begin{align}
    \vec{M} = \chi_M \left( \frac{1}{\mu_0}\vec{B} - \vec{M} \right) =\chi_M \vec{H}. 
\end{align}
If we take a magnetic sphere of radius $R$ and place it in an arbitrary magnetic field, we need to solve for $\nabla \times \vec{H}=\vec{0} $ and $\nabla \cdot \vec{B} =0$ everywhere, which means that $\vec{H}= -\nabla W$ everywhere. Solving for it we have 
\begin{align}
    W_{\rm in} &= \sum_{\ell m} \frac{1+ 2\ell}{1+ 2\ell + \chi_M \ell} C_{\ell m} r^{\ell} Y_{\ell m}, \nonumber \\
     W_{\rm out} &= \sum_{\ell m}  C_{\ell m} \left( r^{\ell} - \frac{\chi_M \ell R^{\ell} }{1+ 2 \ell + \chi_M \ell} \frac{R^{\ell +1}}{r^{\ell+1}} \right) Y_{\ell m} .
\end{align}
where $C_{\ell m}$ are arbitrary that characterize the external magnetic field. From this we have $B_{ {\rm out},\, i }= -\mu_0\partial_i W_{\rm out}$, and the corresponding field strength is calculated through $F_{ij} = \epsilon_{ijk} B^k$. We are interested in the angular components $F_{AB}$ in the outside region, and it is given by 
\begin{align} \label{eqn:FABodd}
    F_{AB} = \epsilon_{ABr} B^r &= -\mu_0 \epsilon_{AB} r^2 \partial_r W \nonumber \\
    & = -\mu_0 \epsilon_{AB} \sum_{\ell m} \ell C_{\ell m}  Y_{\ell m} \left( r^{\ell+1} + \frac{\chi_M (\ell+1) R^{\ell+1} }{1+ 2 \ell + \chi_M \ell} \frac{R^{\ell }}{r^{\ell}} \right).
\end{align}

With this we are now ready to compare it with the solution of Eq.~\eqref{eqn:EFTeom} to find the Love numbers $\lambda_{\ell}^B$ in terms of $\chi_M$. The field strength solved from \eqref{eqn:EFTeom} associated with the boundary condition of external magnetic field is already worked out in \cite{Hui:2020xxx},
\begin{align}
    F_{AB} = \sum_{\ell m} c_{\ell m} \nabla_{[A} \epsilon_{B]}^{\;\; C} \nabla_C Y_{\ell m} \left( r^{\ell+1} + \lambda^B_\ell  \frac{(2\ell -1)!!}{ 4\pi } r^{-\ell}\right) ,
\end{align}
note that $\nabla_{[A} \epsilon_{B]}^{\;\; C} \nabla_C Y_{\ell m}  = \frac{\ell (\ell +1)}{2}\epsilon_{AB} Y_{\ell m}$. Comparing the two expressions we have 
\begin{align}
    \lambda_{\ell}^B = \frac{4\pi \ell  }{ (2\ell -1)!! (1+ 2 \ell + \chi_M \ell)} \chi_M R^{2\ell +1}.
\end{align}
Recall that $\chi_M =-1$ corresponds to superconducting materials and it is natural to expect that \textit{all} isolated materials must have $\chi_{M} \ge -1$. In fact, we will explain in the next subsection that $\chi_M > \frac{1}{1+\chi_E}-1$, as required by subluminality. 

From this expression one can also read off the form of  $a^{(T)}(r)$. From \eqref{eqn:Adecomp} we have 
\begin{align} \label{eqn:FABa}
    F_{AB} = 2 \sum_{\ell m} a^{(T)}_{\ell m}(r) \epsilon_{[B}^{\;\;\; C} \nabla_{A]}  \nabla_C Y_{\ell m}, 
\end{align}
by comparison, the radial dependence of $a^{(T)}$ must be 
\begin{align} \label{eqn:magnetic_ell}
    a^{(T)}(r) \propto  \left( r^{\ell+1} + \lambda^B_\ell  \frac{(2\ell -1)!!}{ 4\pi } r^{-\ell}\right)
\end{align}
This is a gauge invariant quantity which we will focus on for parity odd vector perturbations around a black hole.

\subsection{Causality constraints}
In the above we have been dealing with static problems with constant electromagnetic susceptibilities $\chi_E$ and $\chi_M$. In general, it takes time for the material to respond, therefore the susceptibilities are time dependent, for instance, 
\begin{align}
    \vec{P}(t) = \int d t' \chi_E(t-t') \vec{E}(t') \quad {\rm or} \quad \vec{P}(\omega) = \chi_E(\omega) \vec{E}(\omega). 
\end{align}
In the static limit, we are focusing on the $\omega=0$ part. 
First of all, the constraint from subluminal propagation, $ c^2 = \frac{1}{\epsilon \mu} \le 1$, is obvious, as in linear materials $\epsilon = 1+ \chi_E$, $\mu = 1+ \chi_M$. Therefore we have  
\begin{align}
   (1+\chi_E)(1+\chi_M) \ge 1.
\end{align}
Since we will be interested in the perturbative regime, $\chi_E\ll 1$ and  $\chi_M\ll 1$, we have 
\begin{align} \label{eqn:causalconstr}
    \chi_M + \chi_E \ge 0.
\end{align}
In a more sophisticated consideration of causality, one can derive a set of Kramers-Kronig relations that further constrain $\epsilon$ \footnote{In frequency space, causality requires $\chi(\omega)$ to be analytic in the upper half plane. In terms of $\chi(\omega) = \chi_{ R}(\omega)+ i \chi_{ I}(\omega)$, it can be written as 
\begin{align}
    \chi_R(\omega) = \frac{2}{\pi} \text{P.V.}\int_0^{\infty} d \omega'\, \frac{\omega' \chi_I(\omega')}{\omega'^2-\omega^2}. 
 \end{align}
It is well known that energy dissipation in the material is positively proportional to $\chi_{ I}(\omega)$, which must be positive. Therefore, at zero frequency $\omega=0$, we must have $\epsilon(0) =1+\chi(0) > 1$. 
} and $\mu$. As shown in \cite{Creminelli:2024lhd}, $\epsilon $ is further constrained by 
\begin{align} 
    \epsilon =1 +\chi_E \ge 1 .
\end{align}
Physically, the requirement of having strictly positive $\chi_E$ is simply the fact that polarizations in a material should always align positively with the electric field.

%--------------------------------------------------------------------------------
\section{Vector ($\ell=1$) Love numbers of extremal black holes}
In this section, we examine extremal black holes described by the theory in Eq.~\eqref{eqn:action} placed in a weak external electromagnetic field, focusing on their tidal responses, both electromagnetic and gravitational. Unlike appendix \ref{app:A}, we incorporate gravitational perturbations and highlight the subtleties involved in matching the worldline effective field theory (EFT), arising from the mixing of electromagnetic and gravitational perturbations.

The most general perturbations of the extremal black hole \eqref{eqn:BHmetric} are given by 
\begin{align} \label{eqn:pert}
    h_{\mu\nu} &= \sum_{\ell m} \begin{pmatrix}
        f_t H_0 &  H_1 &  {\cal H}_0 \nabla_A + h_0 \epsilon_A^{\;\;B} \nabla_B \\
         & f_r^{-1} H_2 &  {\cal H}_1 \nabla_A + h_1 \epsilon_A^{\;\;B} \nabla_B \\
            &  &  r^2\left[ K \gamma_{AB} + G \nabla_A \nabla_B + h_2 \epsilon_{(A}^{\;\;\;C} \nabla_{B)} \nabla_C  \right]
    \end{pmatrix} Y^m_{\ell} \nonumber \\
   a_\mu &= \sum_{\ell, m} \left(\begin{array}{c}
a_{0}  \\
a_{r} \\
a^{(L)} \nabla_A   +a^{(T)} \epsilon_A^{\;\; B} \nabla_B  
\end{array}\right) Y^m_{\ell}
\end{align}
for the metric and the gauge field $A_\mu= \bar{A}_\mu + a_{\mu}$. Note that we have suppressed the $(\ell, m)$ indices for the $(t,r)$ dependent functions  $H_0, H_1, H_2 ,\dots, a_r , a^{(L)} $ and $a^{(T)} $. At the linear level, due to spherical symmetry of the background, it is obvious that the parity odd (on the two sphere) sector decouple from parity even sector. We analyze them separately in the following.

\subsection{Parity even sector, electric susceptibility of extremal black hole}
We are now ready to put the extremal black hole in an external electric field. Correspondingly, the parity even part of~\eqref{eqn:pert}, $
\{H_0,\, H_1,\, H_2,\, {\cal H}_0,\, {\cal H}_1,\, K,\, G,\, a_0,\, a_r,\, a^{(L)}\}$, are the relevant perturbations. It is crucial to choose a gauge before the identifying the propagating degrees of freedom. Under a gauge transformation, 
\begin{align}
    \delta h_{\mu\nu} = {\cal L}_{\xi}(\bar{g}_{\mu\nu} + h_{\mu\nu}) , \quad \delta a_{\mu} = {\cal L}_{\xi}(\bar{A}_\mu + a_{\mu}) + \partial_{\mu} \Lambda , 
\end{align}
 with the corresponding diff vector $\xi$ and scalar function $\Lambda$ decomposed into spherical harmonics,
\begin{align}
     \xi &= \xi^{\mu}\partial_{\mu} = \sum_{\ell m} \left[{\cal T}^{\ell m}(t,r) Y^m_{\ell} \partial_t + {\cal R}^{\ell m}(t,r) Y^m_{\ell}\partial_r + \Theta^{\ell m}(t,r) \nabla^AY^m_{\ell} \partial_A \right], \nonumber \\
     \Lambda &= \sum_{\ell m} \Lambda^{\ell m}(t,r) Y^m_{\ell},
 \end{align}
we have the corresponding (leading order) transformations of all the parity even sector functions
\begin{align}
    \delta H_0& = 2 \dot{\cal T} + \frac{f_t'}{f_t}{\cal R}\, , \quad \delta H_1 = f_t {\cal T}' - f_r^{-1} \dot{{\cal R}}\, , \quad \delta H_2 = \frac{f_r'}{f_r} {\cal R} -2{\cal R}'\, , \nonumber \\
    \delta {\cal H}_0 &= f_t {\cal T} - r^2 \dot{\Theta}\, , \quad \delta {\cal H}_1 =-f_r^{-1} {\cal R} - r^2 \Theta'\, ,  \nonumber \\
    \delta K &= -\frac{2}{r} {\cal R} \,, \quad \delta G = -2\Theta\, ,  \nonumber \\ 
    \delta a_0 &= \dot{\bar{A}}_0 {\cal T} +\bar{A}_0 \dot{\cal T} +\bar{A}_0'{\cal R}   + \dot{\Lambda} \, , \quad  \delta a_r = \bar{A}_0{\cal T}' + \Lambda', \quad  \delta a^{(L)} =\bar{A}_0{\cal T}+ \Lambda \,.  
\end{align}
Again, we have suppressed the $(\ell,\, m)$ indices. 

\subsection*{Constant external electric field, dipole perturbation} 
For simplicity, we restrict our self to the case of $\ell =1$ in which the electric field approach a constant asymptotically. In such case the angular metric perturbation $\delta h^{\ell =1}_{AB} $ only depends on the combination ${\cal G} = K-G$. From the above transformations we can adopt the following gauge,
\begin{align} \label{eqn:evengauge}
  H_2 =0\,, \quad   {\cal H}_0 =0 \,, \quad  {\cal G} = K-G =0\,, \quad  a^{(L)} = 0 \, .
\end{align}
We need to stress that for static perturbations, the choice of gauge has to be consistent with the assumptions that all the variables are functions of $r$ only. It is well known that there is an inevitable {\it residual gauge degree of freedom} in the $\ell=1$ sector \cite{Zerilli:1970wzz, Martel:2005ir}. In the above gauge choice, the residual diff is given by
\begin{align}
    {\cal T}^{\ell=1} =0\,, \quad {\cal R}^{\ell=1} = c_0 \sqrt{f_r} \,, \quad \Theta^{\ell=1} = - \frac{c_0}{r} \sqrt{f_r} ,
\end{align}
where $c_0$ is a constant.\footnote{It can be an arbitrary function of time, i.e. $c_0(t)$. For our purposes in analyzing static perturbations, it is a constant.} It preserves the gauge condition~\eqref{eqn:evengauge} but generates the following gauge modes 
\begin{align} \label{eqn:gaugemode}
   \delta H_0 &= c_0  \frac{f_t'}{f_t} \sqrt{f_r} \,, \quad   \delta {\cal H}_1 = c_0 \left( r  \sqrt{f_r}' - \sqrt{f_r} -  \frac{1}{ \sqrt{f_r}} \right)\,, \nonumber \\
  \delta F_{tr} &= - c_0 \partial_r \left(\sqrt{f_r} \bar{F}_{tr} \right)   \,.
\end{align}
Identifying the gauge modes is crucial, as we will see that these modes appear in the solution and must be properly removed.

We are now ready to single out the propagating degrees of freedom associated with the $\ell=1$ perturbations $ \{H_0,\, H_1,\, {\cal H}_1,\,  a_0,\, a_r\}$ in the gauge~\eqref{eqn:evengauge}. The key steps are outlined below, with further technical details provided in the appendix.

It is always natural to consider some gauge invariant quantities for identifying the propagating degrees of freedom. In this problem, one natural choice is the perturbed field strength 
\begin{align}
     \delta F_{tr} = (\dot{a}_r-a_0' )Y^m_{\ell=1}.
\end{align}
Therefore, it motivates us to add the following auxiliary field ${\cal Q}$ in the quadratic action 
\begin{align} \label{eqn:Qtrick}
    \int \rd t \rd r \left [{\cal L}_{(2)}[H_0,\, H_1,\, {\cal H}_1,\,  a_0,\, a_r] - \frac{1}{2} r^2\left(  {\cal Q} -  (\dot{a}_r-a_0' ) \right)^2 \right] ,
\end{align}
which simply gives ${\cal Q} = \dot{a}_r-a_0'$, and obviously leaves the theory unchanged. In the case of pure Einstein-Maxwell theory (i.e. $\alpha_i=0$ in \eqref{eqn:action}), one can solve the equation of motion for $H_1$, $a_0$ and $a_r$ to give
\begin{align} \label{eqn:Qanda}
\mbox{for }\;\alpha_i =0: \quad &  H_1 =\dot{{\cal H}}_1, \nonumber \\
&a_0 = -\left(1-\frac{r_h}{r}\right)^2\left[\frac{1}{2 } \left( r^2{\cal Q}\right)' +\frac{r_h  }{2 \sqrt{2} \kappa } H_0'+\frac{\sqrt{2} r_h }{\kappa  r^2} {\cal H}_1 \right],  \nonumber \\
& a_r = -\left(1-\frac{r_h}{r}\right)^{-2} \left[\frac{1}{2} r^2\dot{\cal Q}+\frac{r_h }{2 \sqrt{2} \kappa } \dot{H}_0 \right].
\end{align}
For pure Einstein-Maxwell, this expression for $\{H_1,\, a_0 ,\, a_r\}$ can be substituted back into the action, so that we are left with the variables $\{{\cal Q},\, H_0 ,\, {\cal H}_1\}$. In this new action, ${\cal H}_1$ can be eliminated, since its equation of motion also becomes algebraic in terms of ${\cal Q},\, H_0 $. Then we are left with a final action in terms of two variables $\{{\cal Q},\, H_0 \}$ and only $\cal{Q}$ is dynamical. 

With the effective field theory, we are working perturbatively in $\alpha$, the above relation should remain true at the zeroth order in $\alpha$. To see this, we apply the same perturbative treatment as in~\cite{DeLuca:2022tkm}, all the function $\{H_0,\, H_1, \dots, {\cal Q} \}$ are expanded perturbatively in the action,
\begin{align}
    H_0 &= H_0^{(0)}+ \alpha H_0^{(1)}, \nonumber \\
    H_1 &= H_1^{(0)}+ \alpha H_1^{(1)},\nonumber \\
     & \dots  \nonumber \\
    {\cal Q} &= {\cal Q}^{(0)}+ \alpha {\cal Q}^{(1)},
\end{align}
and only keep up to ${\cal O}(\alpha)$ in the action. Here $\alpha$ represent collectively the first order expansion in the perturbative parameters $\{\alpha_1,\, \alpha_2,\, \alpha_3\}$ in the action~\eqref{eqn:action}. Now $ \{ H_1^{(1)},\,  a_0^{(1)},\, a_r^{(1)}\}$ become Lagrange multipliers in the action, the equations of motion of $\{H_1^{(1)},\, a_0^{(1)},\, a_r^{(1)}\}$ give the same equations~\eqref{eqn:Qanda} for $H_1^{(0)}$ and relate $a_0^{(0)}$ and $a_r^{(0)}$ to ${\cal Q}^{(0)}$. After substituting these three equations into the action, all the variables $ \{ H_1,\,  a_0,\, a_r\}$ are gone (up to ${\cal O}(\alpha^2)$) and we are left with an action that depend on the rest of the variables,
\begin{align}
    S_{(2)}\left[ H_0^{(0)}, \,H_0^{(1)}, \,{\cal H}_1^{(0)},  \,{\cal H}_1^{(1)},  \,{\cal Q}^{(0)},\,{\cal Q}^{(1)}\right] .
\end{align}
It turns out that ${\cal H}_1^{(1)}$ in this action becomes a Lagrange multiplier, it gives 
\begin{align}
   {\cal H}_1^{(0)} = r^2\frac{ r^2 +r_h^2 }{2 r^2-4 r_h^2} H_0^{(0)}{}'-\frac{r^3 \left(r-2 r_h\right) }{2 \left(r-r_h\right) \left(r^2-2 r_h^2\right)}H_0^{(0)}-\frac{\kappa r_h r^2  }{\sqrt{2} \left(r^2-2 r_h^2\right)} \Big(r^2 {\cal Q}^{(0)}\Big){}'. 
\end{align}
After substituting this into the action, we are left with a final quadratic action that depends on $\{ H_0^{(0)},\,H_0^{(1)}, \,$ ${\cal Q}^{(0)},\,{\cal Q}^{(1)} \}$ only. \footnote{One can also start from the full set of equations of motion of all the variables, followed by repeated substitutions of algebraic equations of motion. The result should be identical to the two equations obtained from the last action. However, the use of action is slightly more systematic.} From this action, the equations of motions for $ H_0^{(0)}$ and ${\cal Q}^{(0)}$ are 
\begin{align} \label{eqn:evenunpert}
   -\frac{r^2 r_h^2}{4 \kappa  \left(r-r_h\right){}^2}\ddot{H}_0^{(0)} +\frac{\left(r-r_h\right){}^2 \left(2 r^2+5 r_h^2\right)}{4 \kappa  \left(r^2-2 r_h^2\right)} H_0^{(0)}{}'' +\frac{(r-r_h) \left(2 r^4-8 r^2 r_h^2+9 r r_h^3-10 r_h^4\right)}{2 \kappa \left(r^2-2 r_h^2\right){}^2}H_0^{(0)}{}'  &\nonumber \\
   -\frac{ \left(r-r_h\right) \left(r^3-r^2 r_h-5 r r_h^2+8 r_h^3\right)}{\kappa  \left(r^2-2 r_h^2\right){}^2} H_0^{(0)} -\frac{r^4 r_h}{2 \sqrt{2}  \left(r-r_h\right)^2}\ddot{\cal Q}^{(0)} +\frac{3 r^2(r-r_h)^2 r_h}{2 \sqrt{2}  \left(r^2-2 r_h^2\right)}{\cal Q}^{(0)}{}'' &\nonumber \\
+\frac{r r_h(r-r_h)  \left(7 r^3-5 r^2 r_h-20 r r_h^2+16 r_h^3\right)}{\sqrt{2} \left(r^2-2 r_h^2\right)^2}{\cal Q}^{(0)}{}' + \frac{\left(r-r_h\right) r_h \left(-r^2 r_h-22 r r_h^2+14 r_h^3+5 r^3\right) }{\sqrt{2} \left(r^2-2 r_h^2\right){}^2}{\cal Q}  & \nonumber \\
  =0& \\
    -\frac{r^4  r_h}{2 \sqrt{2} \kappa  \left(r-r_h\right){}^2}\ddot{H}_0^{(0)}+\frac{3 r^2 \left(r-r_h\right){}^2 r_h}{2 \sqrt{2} \kappa  \left(r^2-2 r_h^2\right)}H_0^{(0)}{}'' -\frac{ rr_h \left(r^4-6 r^3 r_h+9 r^2 r_h^2-4 r_h^4\right)}{\sqrt{2} \kappa   \left(r^2-2 r_h^2\right){}^2}H_0^{(0)}{}' &\nonumber \\
  +\frac{ r_h \left(r^4-6 r^3 r_h+8 r^2 r_h^2-4 r_h^4\right)}{\sqrt{2} \kappa (r^2-2 r_h^2)^2}H_0^{(0)} -\frac{r^6 }{2 \left(r-r_h\right){}^2} \ddot{\cal Q}^{(0)}+ \frac{r^4(r-r_h)^2}{2(r^2-2 r_h^2)}{\cal Q}^{(0)}{}''  &\nonumber \\
    +\frac{r^3  \left(r-r_h\right) \left(2 r^3-r^2 r_h-6 r r_h^2+4 r_h^3\right)}{(r^2-2 r_h^2)^2}{\cal Q}^{(0)}{}' -\frac{r^2 r_h^2 \left(3 r^2-8 r r_h+6 r_h^2\right)}{(r^2-2 r_h^2)^2}{\cal Q}^{(0)} & \nonumber \\
     =0& 
\end{align}
Albeit the appearance of $\ddot{H}_0^{(0)}$ and $\ddot{\cal Q}^{(0)}$, it is easy to verify that there is only one propagating degree of freedom as the kinetic matrix has zero determinant. The first order variables $\{ H_0^{(1)},\,{\cal Q}^{(1)} \}$ satisfies a sourced equation with the same differential operator $D^2$ as in the above equations, schematically 
\begin{align} \label{eqn:perteqneven}
    D^2 \begin{pmatrix}
        H_0^{(1)} \\
        {\cal Q}^{(1)}
    \end{pmatrix} =  {\cal S}\left[H_0^{(0)},\,{\cal Q}^{(0)} \right],
\end{align}
where ${\cal S}$ is the source term in standard perturbation theory that is first order in the coefficients $\alpha_i$. 

For static perturbations, $\dot{H}_0^{(0)}=0$ and $\dot{\cal Q}^{(0)}=0$. There are {\it four} free coefficients in the solution due to the nature of the differential equation. They are fixed by the following procedures:
\begin{itemize}
    \item regularity conditions at the horizon $\Rightarrow$ fix two free coefficients
    \item solutions that are gauge modes of the form~\eqref{eqn:gaugemode} can be removed  $\Rightarrow$ fixes one free coefficients
\end{itemize}
We note that the diff generated profile~\eqref{eqn:gaugemode} is of the form 
\begin{align} \label{eqn:evengaugemode}
    \delta H_0 = c_0 \frac{2r_h}{r^2}, \quad \delta {\cal Q} =  c_0\frac{\sqrt{2} r_h (2 r-3 r_h)}{\kappa  r^4}.
\end{align}
\footnote{Note that due to the ${\cal O}(\alpha)$ correction to the background, there are ${\cal O}(\alpha)$ parts in the gauge modes that we ignored.} Eventually, there is only one free parameter left and the solution that corresponds to an external electric field is given by 
\begin{align}
     H^{(0)}_0 & = C_1 \kappa  r_h (-2 x+3), \nonumber \\
     {\cal Q}^{(0)} & = C_1\frac{ 2x^2-3}{ \sqrt{2} x^2} 
\end{align}
where  $x= \frac{r}{r_h}$  and $C_1$ is the free coefficient that quantifies the strength of the external $\ell=1$ electric field.  Note that ${\cal Q}$ is proportional to the radial component of the electric field.
\begin{align}
    {\cal Q} =\dot{a}_r -a_0' = \frac{E_{r, \ell=1}}{Y^m_{\ell=1}}
\end{align} 
 Comparing with~\eqref{eqn:EMmultipole} we know that asymptotically the dipole radial electric field must be 
\begin{align}
    E_r \simeq V_{\ell m} \left( r^{\ell-1} +\dots   +  \frac{c_\ell}{r^{\ell+2}}  +\dots \right)Y_\ell^m.
\end{align}
The unperturbed solution does not allow any ${\cal O}(r^{-3}) $ fall off tail in ${\cal Q}=E_r$, which implies that the $\ell=1$ electric polarizability of extremal black holes vanish in Einstein-Maxwell theory. It agrees with the well known fact that tidal Love numbers (including scalar, vector and tensor) are vanishing for all charged black holes in Einstein-Maxwell theory. 

The first order perturbations $ H_0^{(1)}$ and ${\cal Q}^{(1)}$ , solved from~\eqref{eqn:perteqneven} after fixing boundary conditions and removing gauge modes, are given by
\begin{align}
 \frac{r_h}{\kappa^3} \alpha H_0^{(1)}    = &C_1 \bigg[-\frac{4}{5} (4 \alpha_1-\alpha_3) x +4 \left(6 \alpha_1-\alpha_3\right) -\frac{16 \alpha_1-4\alpha_3}{5 x} -\frac{4 \left(4 \alpha_1-\alpha_3\right) \ln x}{5 x^2}  \nonumber \\
 &\qquad   - \frac{96 \alpha_1 + 28 \alpha_3 }{3 x^3} \bigg] \nonumber  \\
  \frac{r_h^2}{\kappa^2} \alpha{\cal Q}^{(1)}  =& C_1\bigg[ -\frac{4 \sqrt{2} \left(6 \alpha_1-\alpha_3\right)}{x^2} +\frac{2 \sqrt{2} \big(+300 \alpha_1+25 \alpha_3-(24 \alpha_1-6 \alpha_3) \ln x\big)}{15 x^3}  \nonumber \\ 
&\qquad +\frac{2 \sqrt{2} \big(-352 \alpha_1-47 \alpha_3+ (12 \alpha_1-3 \alpha_3 ) \ln x\big)}{5 x^4}+\frac{16 \sqrt{2} (18 \alpha_1-\alpha_3)}{3 x^5} \nonumber \\ 
&\qquad+\frac{\sqrt{2} (48 \alpha_1+23 \alpha_3)}{x^6} \bigg] .
\end{align}
The presence of $\ln \frac{r}{r_h}$ usually signals a running of the effective coefficient in the point particle EFT~\cite{Goldberger:2009qd,Ivanov:2022hlo}. As the coefficient is directly proportional to $4\alpha_1-\alpha_3$, which is required to be positive by unitarity/WGC, we found an evidence for the following statement:   {\it Positivity and the WGC constrain the RG running of Love numbers}: unitarity demands that logarithmic corrections drive them upwards towards the black hole, ensuring that tidal responses grow rather than diminish toward shorter distances.

\noindent
In other words, the tidal response get weaker away from the black hole.  

Up to here, we just discussed the fall-off and logarithmic correction in the tidal response without really extracting the Love numbers. In order to measure it, one expects that it shows up in the $\frac{1}{r^3}$ tail in ${\cal Q}$ at large $r$. However, the presence of $\frac{\ln (r/r_s)}{r^3}$ in ${\cal Q}^{(1)}$ makes it unclear about the interpretation of the tail $\frac{1}{r^3}$  as $\frac{\ln (r/r_s)}{r^3}$ dominates at large $r$.

In addition, the dipole sector requires careful treatment due to the presence of the gauge mode~\eqref{eqn:evengauge}, given by $\delta H_0 = c_0 \frac{2r_h}{r^2}$ and $\delta {\cal Q} = c_0\frac{\sqrt{2} r_h (2 r-3 r_h)}{\kappa r^4}$, which overlaps with the $\frac{1}{r^3}$ tail in ${\cal Q}^{(1)}$. In the above solution, we eliminated the entire $\frac{1}{r^2}$ component from $H_0^{(1)}$, leaving a residual $\frac{1}{r^3}$ tail in ${\cal Q}^{(1)}$. Alternatively, one could remove the entire $\frac{1}{r^3}$ tail from ${\cal Q}^{(1)}$ instead, resulting in a remaining $\frac{1}{r^2}$ component in $ H_0^{(1)}$. This indicates that the presence of the $\frac{1}{r^3}$ tail in ${\cal Q}^{(1)}$ is, in fact, gauge-dependent and thus ambiguous.

In attempt to resolve the above issues, we consider the following combination 
\begin{align} \label{eqn:QHgaugeinv}
    {\cal Q} - \frac{2r-3r_h}{\sqrt{2}\kappa r^2 } H_0,
\end{align}
which is invariant under the residual gauge transformation~\eqref{eqn:evengaugemode}. For instance, at zeroth order in $\alpha_i$, pure external electric field corresponds to 
\begin{align}
    {\cal Q}^{(0)} - \frac{2r-3r_h}{\sqrt{2}\kappa r^2 } H_0^{(0)} = 3\sqrt{2} C_1 \frac{(r-r_h)^2}{r^2}. 
\end{align}
At first order in $\alpha_i$ we have,
\begin{align} \label{eqn:eveninvsol}
    \alpha {\cal Q}^{(1)} - \frac{2r-3r_h}{\sqrt{2}\kappa r^2 } \alpha H_0^{(1)} &= C_1 \frac{\kappa^2}{r_h^2} \sqrt{2}\bigg[ \frac{ 48 \alpha_1-12 \alpha_3}{15 }+\frac{  78 \alpha_3-432 \alpha_1}{15  x}+\frac{ 228 \alpha_1-42 \alpha_3}{15  x^2} \nonumber \\ 
     &\qquad   +\frac{528 \alpha_1+68 \alpha_3 }{15 x^3}+\frac{-1632 \alpha_1-142 \alpha_3}{15 x^4}+\frac{ 720 \alpha_1-290 \alpha_3}{15  x^5}\nonumber \\ 
     &\qquad  
     +\frac{ 720 \alpha_1+345 \alpha_3 }{15 x^6}  \bigg]. 
\end{align}
The terms with the factor $\ln \frac{r}{r_h} $ are gone in this combination. The strength of the induced tail is determined by the  $x^{-3}$ term \footnote{Here we have chosen the free coefficients such that ${\cal Q}^{(1)}$ does not change the external electric field but with this choice of coefficients, the gauge invariant combination ${\cal Q}^{(1)} - \frac{2r-3r_h}{\sqrt{2}\kappa r^2 } H_0^{(1)}$ contains a constant term, which appear to correct the external field. However, one can choose the coefficients such that there is no ${\cal O}(x^0)$ term in ${\cal Q}^{(1)} - \frac{2r-3r_h}{\sqrt{2}\kappa r^2 } H_0^{(1)}$ but now there is an ${\cal O}(x^0)$ term in  ${\cal Q}^{(1)}$. Fortunately, both choices {\it do not} affect the coefficient of the tail  ${\cal O}(x^{-3})$ therefore the Love number is insensitive to this choice. }, which is 
\begin{align}
  C_1 \frac{\kappa^2}{r_h^2} \sqrt{2}  \frac{528 \alpha_1+68 \alpha_3 }{15 x^3}.
\end{align}
In the following we pretend that this is the correct combination for the electric field strength when we compare it with the worldline EFT. By comparison with the electric field~\eqref{eqn:radialE} for $\ell=1$, one concludes that perturbatively, 
\begin{align} \label{eqn:chiE}
     \chi_E \approx \frac{\kappa^2}{r_h^2} \frac{264 \alpha_1+34 \alpha_3}{15} >0
\end{align}
which implies that in order to have positive electric deformability, we should demand
\begin{align} \label{eqn:alphaEG}
     \alpha_1 > - \frac{17}{132} \alpha_3.
\end{align}
This constraint is weaker than the unitarity constraint $\alpha_1 >\frac{1}{4}|\alpha_3|$. In other words,  \textit{unitarity guarantees the positivity of $\ell=1$ electric polarizability}. 

We emphasize that this is only a preliminary comparison with the worldline EFT, as the gauge issue and the correct combination of the electric field strength should also be handled in the worldline EFT. A more complete treatment in the worldline EFT, incorporating both gravitational and electromagnetic perturbations, is required for strongly charged objects.

Before moving to the parity odd sector, we must highlight the difference between~\eqref{eqn:alphaEG} in the above and~\eqref{eqn:alphaEpure} in section~\ref{sec:evenpureEM}, where gravitational perturbations are ignored. Also in the electromagnetic part of the worldline EFT calculations, mixing with gravitational perturbations is ignored. However, for an extremal black hole, both the electromagnetic and gravitational perturbations are equally important due to mixing. The computation in this section exactly demonstrates this point by showing that the ${\cal O}(\alpha_i)$ perturbed field profile is corrected when this graviton-photon mixing is taken into account.

\subsection{Parity odd sector, magnetic susceptibility}
Next we consider the parity odd sector. The relevant perturbations are the parity odd part of of~\eqref{eqn:pert}, given by $\{h_0,\, h_1,\, h_2,\, a^{(T)}\}$. By construction, $a^{(T)}$ is gauge invariant while the rest transform under a parity odd diff vector,
\begin{align*}
     \xi_{\rm odd} =\xi_{\rm odd}^{\mu}\partial_{\mu} = \sum_{\ell m} \left[ \xi_{V}^{\ell m} r^{-2} \epsilon^{AB}\nabla_B Y^m_{\ell}  \partial_A \right] ,
\end{align*}
as
\begin{align} \label{eqn:oddgaugetran}
    \delta h_0 = \dot{\xi}_{V}, \quad \delta h_1 = \xi_V' - \frac{2}{r} \xi_V, \quad \delta h_2 = -2 \xi_V.
\end{align}
Again, the $(\ell,\, m)$-indexes are suppressed. Therefore, one can work with the convenient gauge choice $h_2=0$ for $\ell \ge 2$, as this is a complete gauge fixing. 

From now on, we focus on $\ell =1$ in the parity odd sector. In other words, we are looking for the response of a black hole in an external constant magnetic field. In this case, $\epsilon_{(A}^{\;\;\;C} \nabla_{B)} \nabla_C  Y^m_{\ell=1} =0$, therefore $h_2$ is irrelevant and we only have the functions $h_0$ and $h_1$. From the gauge transformations~\eqref{eqn:oddgaugetran} it might appear that either $h_0$ or $h_1$ can be set to zero. However, extra caution is needed for the specific problem we are considering. As we focus on static perturbations, there is a time-like Killing vector orthogonal to $\partial_r$ for both the background and the linear solution, implying that the perturbations depend only on $r$ in this gauge choice. To preserve this choice of gauge, there is a limited set of diff vectors $\xi_V$ available. We work with the convenient gauge choice
\begin{align}
    h_1^{\ell=1}=0,
\end{align} 
since from~\eqref{eqn:oddgaugetran} one can remove  $h_1^{\ell=1}$ by choosing a $r$-dependent $\xi_V$. \footnote{If one wishes to gauge away $h_0(r)$, which depends on $r$ only in the chosen gauge, the gauge transformation has to be $\xi_V= -t h_0(r)$ due to~\eqref{eqn:oddgaugetran}. Such a gauge change would introduce time dependence in $h_1$ and would break the assumption that the time-like killing vector is orthogonal to the radial direction and we can no longer assume that the perturbations are purely functions of $r$. While the analysis could still be carried out, it would introduce unnecessary complications.} 

\subsection*{Residual gauge mode} 
The $\ell=1$ sector cannot be completely gauge fixed. In this gauge, there exist a residual diff that does not introduce time dependence, given by
\begin{align}
    \xi_V = c_0 t r^2 ,
\end{align}
that generates the gauge mode 
\begin{align}
    h_0^{\ell =1} = c_0r^2.
\end{align}
In other words, any $r^2$ term in $h_0^{\ell =1}$ can be removed using the residual diff.

\subsection*{Rotation mode} 
In addition to the residual gauge freedom, it is well known that the $\ell=1$ sector contains a solution that corresponds to a small rotation of the black hole background \cite{Martel:2005ir}. This mode takes the form 
\begin{align}
    h_0= \frac{J}{r}
\end{align}
for some constant $J$ that is related to angular momentum.

\subsection*{Physical solutions for $h_0$ and $a^{(T)}$ for $\ell=1$} 
Now we are ready to discuss the $\ell =1$ perturbations for an extremal black hole. Carrying out the usual first order perturbative expansion in $\alpha_i$, 
\begin{align}
    h_0 &= h_0^{(0)} + \alpha h_0^{(1)} + \dots, \nonumber \\
    a^{(T)}&= a^{(0)} + \alpha  a^{(1)}+ \dots,
\end{align}
the equations of motion for the zeroth order (in $\alpha_i$) perturbations are 
\begin{align} \label{eqn:odd_linear}
    \frac{\sqrt{2} r_h}{r^2} \partial_ra^{(0)}{}+\frac{1}{2 \kappa } \partial_r^2h_0^{(0)}{}-\frac{1}{\kappa  r^2} h_0^{(0)} &=0 \nonumber \\
-\frac{r^2}{ (r-r_h)^2 } \ddot{a}^{(0)}+\frac{ (r-r_h)^2 }{r^2}\partial_r^2a^{(0)}{}+2\frac{  r_h (r-r_h) }{r^3}\partial_ra^{(0)}{}-\frac{2  }{r^2} a^{(0)} 
+\frac{\sqrt{2}   r_h }{\kappa  r^2}\partial_r h_0^{(0)}{}-\frac{2 \sqrt{2}   r_h }{\kappa  r^3}h_0^{(0)} &=0
\end{align}
We look for solutions that describe static external magnetic field. After imposing regularity at the horizon, eliminating the gauge mode, and removing the mode associated with small rotation, the solution is given by:
\begin{align}
  \kappa^{-1}  h_0^{(0)} &= C_a 2\sqrt{2} x, \nonumber \\
    a^{(0)}{} &= C_a(x^2-3),
\end{align}
where $x=\frac{r}{r_h}$ . This is exactly an $\ell=1$ external magnetic field in the asymptotic region and $C_a$ quantifies the strength of it. The first order in $\alpha_i$  perturbed parts $\{h_0^{(1)}, a^{(1)}\}$ satisfy a set of sourced differential equations with the same differential operator as~\eqref{eqn:odd_linear} with source terms in terms of $\{h_0^{(0)}, a^{(0)}\}$, schematically,
\begin{align}
    \tilde{D}^2 \begin{pmatrix}
        h_0^{(1)} \\a^{(1)} 
    \end{pmatrix} =  \tilde{{\cal S}}[h_0^{(0)}, a^{(0)} ]
\end{align}
The next-order corrections $\{h_0^{(1)}, a^{(1)}\}$ must satisfy the same set of conditions as the leading-order solution, including regularity at the horizon and the absence of gauge and rotational modes. The results are 
\begin{align}
  \kappa^{-1}   \alpha h_0^{(1)} &= C_a\frac{\kappa^2}{r_h^2} \Bigg[ \frac{8 \sqrt{2}  }{3 }  (6 \alpha_1-\alpha_3)x 
    -\frac{8 \sqrt{2}   }{5  }(6 \alpha_1-4 \alpha_2+3 \alpha_3) \frac{1}{x^2} \nonumber \\
 & \qquad  \qquad \qquad    +\frac{4 \sqrt{2}  }{75 }(264 \alpha_1-96 \alpha_2-103 \alpha_3) \frac{1}{x^3} + {\cal O}(x^{-4}) \Bigg], \nonumber \\
   \alpha a^{(1)} &=  C_a\frac{\kappa^2}{r_h^2} \Bigg[8 (\alpha_3-6 \alpha_1)
      +\frac{32}{5 }  (2\alpha_2+\alpha_3-3 \alpha_1)\frac{1}{x}
   +\frac{4  }{15  } (132 \alpha_1-48\alpha_2-49 \alpha_3) \frac{1}{x^2} \nonumber \\   
  &\qquad \qquad \qquad     + \frac{16 }{25} (8 \alpha_1-32\alpha_2+9 \alpha_3) \frac{1}{x^3} + {\cal O}(x^{-4}) \Bigg].
\end{align}
Compared to~\eqref{eqn:FABodd}, the induced magnetic response is $C_a\frac{\kappa^2}{r_h^2}\frac{32}{5 } (2\alpha_2+\alpha_3-3 \alpha_1)\frac{1}{x}+ \dots $ with
\begin{align} 
    \chi_M \approx \frac{\kappa^2}{r_h^2}\frac{48}{5 } (2\alpha_2+\alpha_3-3 \alpha_1).
\end{align}

In the absence of logarithmic corrections, comparison with a simple EFT calculation in the $\ell=1$ sector should be safe, meaning that gravitational mixing in the EFT can be ignored. Furthermore, one can combine with the result from the parity even sector~\eqref{eqn:chiE} and compare with the causality constraint~\eqref{eqn:causalconstr} $\chi_E + \chi _M >0$ (in the perturbative region). We have 
\begin{align}
    \chi_E+ \chi_M \approx \frac{\kappa ^2}{r_h^2} \frac{2  (- 84  \alpha_1 +144 \alpha_2 +89 \alpha_3)}{15 } >0. 
\end{align}
This {\it preliminary} constraint on the EFT coefficients are different from all existing results. Again, as we mentioned in the previous section, due to the existence of the logarithmic term and mixing with gravity in the extremal limit, the EFT matching should be carried out with more careful analysis. We will leave this for future work.  

\section{Magnetic and parity odd tensor tidal responses for $\ell \ge 2$}
In the $\ell=1$ sector, the spin-2 gravitational perturbations are non-dynamical and contain residual gauge degrees of freedom. They become dynamical only for $\ell \ge 2$, where the analysis differs substantially: although no residual gauge freedom remains, the mixing between the two dynamical modes induced by the black hole’s charge introduces an additional layer of complexity.

In this section, we focus on the parity-odd (magnetic) sector of the perturbations where only the functions $\{ h_0, \, h_1, \, h_2, \, a^{(T)}  \}$ in~\eqref{eqn:pert} are relevant  \footnote{The analysis  for $\ell \ge 2$ in pure Einstein-Maxwell theory has been done in \cite{Pereniguez:2021xcj}. However, it becomes a technically complicated task to solve for the parity even part perturbatively in the EFT~\eqref{eqn:action}. We leave this part for future exploration.},
\begin{align} \label{eqn:pertodd}
    h^{\rm odd}_{\mu\nu} &= \sum_{\ell m} \begin{pmatrix}
        0 & 0 &   h_0^{(\ell m)}(t,\,r) \epsilon_A^{\;\;C} \nabla_C \\
       0  & 0 &   h_1^{(\ell m)}(t,\,r) \epsilon_A^{\;\;C} \nabla_C \\
        h_0^{(\ell m)}(t,\,r) \epsilon_B^{\;\;C} \nabla_C    &  h_1^{(\ell m)}(t,\,r) \epsilon_B^{\;\;C} \nabla_C  &  r^2 h_2^{(\ell m)}(t,\,r) \epsilon_{(A}^{\;\;\;C} \nabla_{B)} \nabla_C 
    \end{pmatrix} Y^m_{\ell}, \nonumber \\
   a^{\rm odd}_\mu &= \sum_{\ell, m} \left(\begin{array}{c}
0  \\
0 \\
  a^{(T)}_{(\ell m)}(t,\,r) \epsilon_A^{\;\; B} \nabla_B  
\end{array}\right) Y^m_{\ell} . 
\end{align}
Starting from now on, we will suppress the $(\ell, m)$ indices in the functions. In this sector, the most convenient gauge choice is the Regge-Wheeler gauge 
\begin{align}
    h_2 =0.
\end{align}
Therefore the quadratic action is a functional of three functions only, i.e. $S_{(2)}[ h_0, \, h_1, \, a^{(T)}]$. 

To simplify the analysis, we use a similar trick as in~\eqref{eqn:Qtrick} and~\cite{DeLuca:2022tkm}. An auxiliary variable ${\cal Q}(t,\,r)$ is introduced in the action as 
\begin{align}  \label{eqn:lge2Qaction}
    \tilde{S}_{(2)} = \int \rd t \rd r \left [ {\cal L}_{(2)}[ h_0, \, h_1, \, a^{(T)}]  - \frac{1}{4\kappa^2} \left( \frac{\sqrt{2} \mu}{r} {\cal Q} +\frac{2\sqrt{2} r_h}{r} a^{(T)} + h_0' - \frac{2}{r}h_0 - \dot{h}_1 \right)^2 \right ],
\end{align}
where   $\mu^2 = \ell(\ell+1) -2 $. Written in this form, the action is clearly equivalent to the original one once the algebraic equation of motion for ${\cal Q}$ is imposed. In the $\alpha_i=0$ limit, this action gives rise to algebraic equations of motion for $h_0$ and $h_1$,
\begin{align} \label{eqn:lge2Qh}
    h_0 &= - \frac{\sqrt{2} (r-r_h)^2}{\mu  r^2} \left(  r {\cal Q }  \right )' , \nonumber \\
    h_1 &=  -\frac{\sqrt{2} r^3}{\mu (r-r_h)^2 } \dot{\cal Q}. 
\end{align}
Once the four derivative corrections are introduced ($\alpha_i \neq 0$), we work perturbatively in the action up to first order in $\alpha_i$ by expanding the functions,
\begin{align}
    h_0 &= h_0^{(0)}+ \alpha h_0^{(1)} + \dots , \nonumber \\
    h_1 &= h_1^{(0)}+ \alpha h_1^{(1)} \dots ,\nonumber \\
    a^{(T)} &= a^{(0)}+ \alpha a^{(1)} \dots ,\nonumber \\
    {\cal Q} &= {\cal Q}^{(0)}+ \alpha {\cal Q}^{(1)} ,
\end{align}
where $\alpha$ here collectively denote $\{\alpha_i\}$. In this way, the equations of motion from the action give the same relation~\eqref{eqn:lge2Qh} for both the zeroth order functions $\{h_0^{(0)},\,h_1^{(0)},\, {\cal Q}^{(0)} \} $ and the first order functions $\{h_0^{(1)},\,h_1^{(1)},\, {\cal Q}^{(1)}\}$. This procedure yields the same perturbative equations as directly using the equations of motion, but working at the level of the action is far more systematic. Therefore, the algebraic expressions for $\{h_0^{(0)},\,h_1^{(0)},\}$ and $\{h_0^{(0)},\,h_1^{(0)},\}$ in terms of ${\cal Q}^{(0)}$, ${\cal Q}^{(1)}$ can be substituted into the action, resulting in an action , 
 \begin{align}
     S_{(2)}[a^{(0)}, \, {\cal Q}^{(0)} ,\, a^{(1)} ,\, {\cal Q}^{(1)}],
 \end{align}
 that is a functional of $\{a^{(0)}, \, {\cal Q}^{(0)} ,\, a^{(1)} ,\, {\cal Q}^{(1)}\}$ only. The equations of motion from this action are 
\begin{align} \label{eqn:lge2}
    \frac{r^2 }{\left(r-r_h\right){}^2} \ddot{\cal Q}^{(0)}-\frac{ \left(r-r_h\right){}^2}{r^2}{\cal Q}^{(0)}{}'' &+\frac{2 \left(r-r_h\right) r_h}{r^3} {\cal Q}^{(0)}{}'  \nonumber \\ 
    & +\frac{\left(\mu ^2+2\right) r^2 -6 r r_h+4 r_h^2}{r^4} {\cal Q}^{(0)}+ \frac{2  \kappa  \mu   r_h}{r^3} a^{(0)}= 0, \nonumber \\
   \frac{r^2 }{\left(r-r_h\right){}^2} \ddot{a}^{(0)}-\frac{ \left(r-r_h\right){}^2}{r^2} a^{(0)}{}'' & +\frac{2 \left(r-r_h\right) r_h}{r^3} a^{(0)}{}'  \nonumber \\ 
   & +  \frac{ \left(\mu ^2+2\right) r^2+4 r_h^2 }{  r^4} a^{(0)} +\frac{2 \mu  r_h }{\kappa  r^3} {\cal Q}^{(0)} = 0 ,
\end{align}
and the first order functions $\{ a^{(1)} ,\, {\cal Q}^{(1)}\}$ satisfy a set of sourced differential equations with the {\it same} differential operator, schematically
\begin{align}
       D^2 \begin{pmatrix}
        {\cal Q}^{(1)} \\
        a^{(1)}
    \end{pmatrix} =  {\cal S}\left[a^{(0)},\,{\cal Q}^{(0)} \right]. 
\end{align}
We omit many of the explicit source terms, as their detailed form offers little additional physical insight. What is important, and already evident in this set of equations, is the mixing between the gravitational and electromagnetic perturbations induced by the black hole’s charge. One can work with a diagonal basis by introducing~\cite{Chandrasekhar:1985kt}
\begin{align} \label{eqn:Qadiag}
    \begin{pmatrix}
        {\cal Q} \\
        a
    \end{pmatrix}  
    = \begin{pmatrix}
        \kappa \cos \delta & \kappa \sin \delta \\
        -\sin \delta & \cos \delta 
    \end{pmatrix} \begin{pmatrix}
        U \\
        V
    \end{pmatrix}  , \qquad    \cos \delta = \sqrt{\frac{3}{2 \sqrt{4 \mu ^2+9}}+\frac{1}{2}}.
\end{align}
Again, $\mu^2 = \ell (\ell +1) -2 $. The zeroth order functions $\{U^{(0)}, \, V^{(0)} \}$ satisfy the following equations of motion,
\begin{align}
    \frac{r^2 }{\left(r-r_h\right){}^2} \ddot{U}^{(0)}-\frac{ \left(r-r_h\right){}^2}{r^2} U^{(0)}{}'' &+\frac{2 \left(r-r_h\right) r_h}{r^3} U^{(0)}{}'  \nonumber \\ 
    & +\frac{ \left(\mu ^2+2\right) r^2-\left(\sqrt{4 \mu ^2+9} {\color{magenta}+ }3\right) r r_h +4 r_h^2}{r^4} U^{(0)} = 0, \nonumber \\
   \frac{r^2 }{\left(r-r_h\right){}^2} \ddot{V}^{(0)}-\frac{ \left(r-r_h\right){}^2}{r^2} V^{(0)}{}'' & +\frac{2 \left(r-r_h\right) r_h}{r^3} V^{(0)}{}'  \nonumber \\ 
   & +\frac{ \left(\mu ^2+2\right) r^2-\left(\sqrt{4 \mu ^2+9} {\color{magenta}- }3\right) r r_h +4 r_h^2}{r^4} V^{(0)} = 0 ,
\end{align}
where the colored part highlight the only difference in these two equations. Similarly, the first order functions $\{  U^{(1)} ,\, V^{(1)}\}$ satisfy a sourced differential equation with the same differential operator. At zeroth order, the most general static solutions are 
\begin{align}
    U^{(0)}  &= C_U  \frac{\ell (\ell -1)  (2 \ell-1) r^3+3 (\ell-1) (2 \ell-1) r^2 r_h+9 (\ell-1) r r_h^2+6 r_h^3}{r} (r-r_h)^{\ell-1} \nonumber \\
   &\qquad   + D_U\frac{ (\ell+1) r-2 r_h }{r (r-r_h)^{\ell} },  \nonumber \\
   V^{(0)} & =   C_V \frac{ (r-r_h)^{\ell+1} (l r+2 r_h)}{r} \nonumber \\
   &\qquad + D_V  \frac{ (\ell+1) (\ell+2) (2 \ell+3) r^3-3 (\ell+2) (2 \ell+3) r^2 r_h+9 (\ell+2) r r_h^2-6 r_h^3}{r (r-r_h)^{\ell+2}} . 
\end{align}
By imposing regularity at the horizon, we set $D_U = D_V = 0$, and the growing part correspond to external tidal/magnetic fields.

Before presenting the next-order solutions in $\alpha_i$, it is useful to clarify the definition of an external tensor tidal field and its associated response in the presence of mixing. In particular, we must specify what we mean by a ``pure'' external gravitational field versus a ``pure'' external magnetic field in the asymptotic region $r \to \infty$. These can be defined by the leading asymptotic behavior of the fields:
\begin{align*}
    \text{external odd gravitational field only:} \qquad  & {\cal Q} \sim r^{\ell+1}, \qquad 
      a \;\text{growing no faster than}\; r^\ell , \\
    \text{external magnetic field only:} \qquad  & a \sim r^{\ell+1}, \qquad
      {\cal Q} \;\text{growing no faster than}\; r^\ell .
\end{align*}
Using~\eqref{eqn:Qadiag}, they translate into the choice of coefficients $C_U$ and $C_V$. These asymptotic conditions distinguish the two types of external perturbations and define the corresponding tidal responses. We should mention that it is impossible to have a consistent solution such that it is ``purely" gravitational with ${\cal Q}\neq 0 ,\,a=0$, due to mixing.  

Obviously, due to mixing, there exist mixed type of responses, such as gravitationally induced electromagnetic response and electromagnetically induced gravitational response.

\subsection{ $\ell =2$ and $\ell =3$ responses }
At first order in $\alpha_i$, the general $\ell$ solution can be obtained by the convolving the Green's function with the source. However, the expression is messy therefore we choose to just display the answer for $\ell=2$ and $\ell=3$ for physical discussion. 

\noindent
\underline{$\ell=2$:} 

\noindent
To present the results for $\ell=2$, we separate them into two cases: external  odd gravitational field only, or external magnetic field only. The most general solution is just a linear combination of the two. 

The $\ell=2$, external odd gravitational field solution at the zeroth order is 
\begin{align}
    {\cal Q}^{(0)}& =  \kappa  C_{\cal Q}    \left(x^3- \frac{1}{x}\right), \nonumber \\
    a^{(0)}& =  -C_{\cal Q} (x^2-1)
\end{align}
where $x=\frac{r}{r_h}$. In the first order in $\alpha_i$, the regular solution is 
\begin{align}
    \alpha {\cal Q}^{(1)}& = \frac{C_{\cal Q} \kappa^3}{r_h^2}\bigg[ -\frac{4}{3}   (6 \alpha_1-\alpha_3) x  -\frac{24 \alpha_1-62 \alpha_3}{x} \nonumber \\
     & \quad\quad\qquad  -\frac{8 \left(56 \alpha_1-128 \alpha_2-341 \alpha _3+ {\color{purple}360 \alpha_3} \ln x \right)}{25 x^2}+ {\cal O}\left(\frac{1}{x^3} \right) \bigg], \nonumber \\
   \alpha  a^{(1)}& = \frac{C_{\cal Q} \kappa^2}{r_h^2}  \bigg[ -\frac{4}{3}  \left(6 \alpha_1-\alpha_3\right) x^2+4 \left(6 \alpha_1-7 \alpha_3\right)+\frac{32 \alpha_3}{x} \nonumber \\
     & \quad\quad\qquad + 8\frac{ -1170 \alpha_1+852 \alpha_2+1621 \alpha_3+ 90 {\color{orange}\left(8 \alpha_1 - 8 \alpha_2- 9 \alpha_3 \right)} \ln x}{75 x^2} + {\cal O}\left(\frac{1}{x^3} \right) \bigg]. 
\end{align}
It is obvious from this expression that there is running Love numbers starting from exactly the power  ${\cal O}(x^{-2})$ of the response. 

For the external magnetic field only setup, the zeroth order solution is 
\begin{align}
    {\cal Q}^{(0)}& = -  \kappa C_{a}  \left(x^2- 1\right), \nonumber \\
    a^{(0)}& =  C_{a}   \left( x^3-\frac{3}{2}x^2 +\frac{3}{2} - \frac{1}{x} \right),
\end{align}
and in first order in $\alpha_i$,
\begin{align}
    \alpha   {\cal Q}^{(1)}& =\frac{C_{a} \kappa^3}{r_h^2}\bigg[ \left(2\alpha_3-12 \alpha _1\right) x+  \left(24 \alpha_1- 4\alpha _3\right)+\frac{12  \alpha _1+4\alpha _3}{x}  \nonumber \\
     & \quad\quad\qquad  +2\frac{-7380 \alpha _1+5568 \alpha _2+5069 \alpha _3+ 360  {\color{orange}\left(8 \alpha _1-8 \alpha _2 - 9\alpha _3 \right)} \ln x}{75 x^2} +  {\cal O}\left(\frac{1}{x^3} \right) \bigg], \nonumber \\
   \alpha   a^{(1)}& =  \frac{C_{a} \kappa^2}{r_h^2}  \bigg[ \frac{7 \alpha _3-48 \alpha _1}{5}  x^2 +\frac{3}{5} \left(56 \alpha _1+11 \alpha _3\right) -\frac{2 \left(192 \alpha _1-144 \alpha _2-11 \alpha _3\right)}{3 x}  \nonumber \\
       & \quad\quad\qquad   + 8\frac{ -10967 \alpha _1+6638 \alpha _2+3563 \alpha _3+ 630 {\color{cyan} ( 12 \alpha _1-8 \alpha _2-3 \alpha _3 )}\ln x }{175 x^2} +  {\cal O}\left(\frac{1}{x^3} \right) \bigg].
\end{align}
In the presence of logarithmic correction, one can change the scale in the logarithmic term as $\ln \frac{r}{\sigma   r_h}$ for some number $\sigma$ such that the coefficients of the power law changes correspondingly. Also, in a shift of the radial coordinate, for instance $r \to r -r_h$, the logarithmic term remains while the power law coefficients mix. Only coefficient of the $\ln x$ term is meaningful as they are gauge invariant.  The {\color{purple} purple} colored part is the gravitationally induced tidal response and the {\color{cyan} cyan} colored part is the magnetically induced EM response. The {\color{orange} orange} colored parts tell us about the cross gravito-EM responses. They are {\it exactly the same} as they come from the cross induction, i.e. gravitaionally induced magnetic multipole and magnetically induced gravitaional multipole. It would be clearer from the worldline EFT point of view which we will discuss later.\\

\noindent
\underline{$\ell=3$ pure external {\it gravitational} field:}  

\noindent
The zeroth order solution is 
\begin{align}
        {\cal Q}^{(0)}& =  \kappa C_{\cal Q} (x-1)^2  \left( x^2+\frac{1}{3}x+\frac{1}{3}+\frac{1}{3x}\right), \nonumber \\
    a^{(0)}& = C_{\cal Q}\sqrt{\frac{2}{5}}  \left( -\frac{5}{3} x^3+ \frac{8}{3} x^2 -\frac{4}{3} x+ \frac{1}{3x}\right),
\end{align}
For the ${\cal O}(\alpha_i)$ solution, we only show the logarithmic term at the corresponding order $x^{-\ell}$ of the response, as they are gauge invariant and free of other ambiguities such as choice of scale  \footnote{In the top down picture, when the gravitational EFT (with cutoff $\Lambda$) is further reduced to the worldline EFT for a single black hole, one chooses a radius $R > r_h > 1/\Lambda$, in which all internal microscopic structures within this radius is effectively described by the effective worldline operators. Specifying this radius $R$ at different locations is analogous to picking a renormalization scale. In the presence of logarithmic terms in the bulk gravitational EFT solution, as $\ln \frac{r}{ r_h } = \ln \frac{r}{ R }+ \ln \frac{R}{ r_h }$, when we match the bulk solution to the worldline EFT at the scale $R$, this constant piece gets absorbed into the definition of the finite, non logarithmic tidal response, which is induced by the worldline Wilson coefficient (the Love number). }. 
\begin{align}
\alpha {\cal Q}^{(1)} & \supset  \frac{C_{\cal Q} \kappa^3}{r_h^2} \frac{64}{21}  (116 \alpha_1-128 \alpha_2-195 \alpha_3)\frac{ \ln x }{ x^3} ,\nonumber \\ 
  \alpha a^{(1)} &\supset  \frac{C_{\cal Q} \kappa^2}{r_h^2}  {\color{orange} \sqrt{\frac{2}{5}}  \frac{64}{21} (500 \alpha_1-392 \alpha_2-351 \alpha_3) }\frac{  \ln x }{ x^3} .
\end{align}

\noindent
\underline{$\ell=3$ pure external {\it magnetic} field:}  

\noindent
The zeroth order solution is 
\begin{align}
    {\cal Q}^{(0)}& =  \kappa C_{a}  \sqrt{\frac{2}{5}}\left(\frac{-5}{3} x^3+ \frac{8}{3} x^2- \frac{4}{3} + \frac{1}{3 x}\right)  , \nonumber \\
     a^{(0)}& = C_{a} (x-1)^2 \left(  x^2- \frac{10}{15} x-\frac{1}{15}+\frac{8}{15 x} \right).
\end{align}
At ${\cal O}(\alpha_i)$, the logarithmic running terms are
\begin{align}
      \alpha  {\cal Q}^{(1)}  & \supset \frac{C_{a} \kappa^2}{r_h^2}   {\color{orange}\sqrt{\frac{2}{5}}  \frac{64}{21} (500 \alpha_1-392 \alpha_2-351 \alpha_3)} \frac{  \ln x }{ x^3} ,  \nonumber \\
  \alpha a^{(1)} &\supset  \frac{C_{a} \kappa^2}{r_h^2}  \frac{256}{105} \left(892 \alpha _1-544 \alpha _2-339 \alpha _3\right) \frac{ \ln x}{ x^3}
\end{align}
Again, the orange colored part highlights the cross responses, the magnetically induced gravitational response and the gravitationally induced magnetic response are the same. In the following, we will see that this is expected, from the point of view of the worldline effective field theory.

%%%%%%%%%%%%%%%%%%%%%%%%%%%%%%%%%%%%%%%%%%%

\subsection{Worldline EFT operators with gravito-EM mixing}
For charged compact objects, gravitational and electromagnetic perturbations inevitably mix at the linear level, leading to cross-responses. A pure gravitational tidal field, for instance, can induce an electromagnetic response, and an external electromagnetic field can generate a gravitational tidal deformation. They are intuitively natural. These mixed responses reflect the coupled nature of perturbations around a charged black hole and must be accounted for in a complete worldline EFT description.~\cite{Porto:2016pyg} 

From the bottom up worldline EFT point of view, it is straightforward to construct worldline operators that describe tidal and electromagnetic deformations. For example, the operators
\begin{align}
    \int d\tau \, {\cal E}_{\langle L\rangle } {\cal E}^{\langle L\rangle },  \quad \int d\tau  \, {\cal B}_{\langle L\rangle } {\cal B}^{\langle L\rangle },  \quad  \int d\tau \, E^{\rm EM}_{\langle L\rangle } E_{\rm EM}^{\langle L\rangle },  \quad \int d\tau  \, B^{\rm EM}_{\langle L\rangle }B_{\rm EM}^{\langle L\rangle },  
\end{align}
describe the well-studied linear tensor and vector deformations~\cite{Goldberger:2004jt,Kol:2011vg,Porto:2016pyg,Hui:2020xxx}. Here, the electric and magnetic components of the Weyl tensor on the worldline are defined in the usual way:
\begin{align}
   {\cal E}_{\mu\nu} =  W_{\mu \rho \nu \sigma } u^{\rho }u^{\sigma} , \quad   {\cal B}_{\mu\nu} =  \tilde{W}_{\mu \rho \nu \sigma } u^{\rho }u^{\sigma}.
\end{align}
Also, their higher multipole parts ($L >2 $) are captured by the following tensors:
\begin{align}
    {\cal E}_{\langle L\rangle } = \nabla_{ \langle \mu_1} \dots   \nabla_{\mu_{L-2}} {\cal E}_{\mu_{L-1} \mu_{L} \rangle},  \quad  {\cal B}_{\langle L \rangle } = \nabla_{\langle \mu_1} \dots   \nabla_{\mu_{L-2}} {\cal B}_{\mu_{L-1} \mu_{L} \rangle },
\end{align}
where $\langle \ \;\rangle $ denotes the symmetric trace free part. In a similar manner, the higher multipole moments of the electric and magnetic fields on the worldline are captured by:
\begin{align}
   E^{\rm EM}_{\langle L\rangle } = \nabla_{ \langle \mu_1} \dots   \nabla_{\mu_{L-1}} E_{\mu_{L} \rangle},  \quad  B^{\rm EM}_{\langle L \rangle } = \nabla_{\langle \mu_1} \dots   \nabla_{\mu_{L-1}} B_{ \mu_{L} \rangle }. 
\end{align}

In the presence of gravito-electromagnetic mixing, we can introduce the following worldline operators:
\begin{align}
    \int d\tau \, {\cal E}_{\langle L\rangle }  E_{\rm EM}^{\langle L\rangle },  \quad \int d\tau  \,  {\cal B}_{\langle L\rangle } B_{\rm EM}^{\langle L\rangle } ,
\end{align}
\footnote{
In addition to the parity-conserving operators, a set of parity-violating operators can be constructed, which involve E-B type mixing:
\begin{align}
    \mathcal{E}_{\langle L\rangle }\mathcal{B}^{\langle L\rangle }, \quad E_{\langle L\rangle }B^{\langle L\rangle }, \quad \mathcal{E}_{\langle L\rangle }B^{\langle L\rangle }, \quad \mathcal{E}_{\langle L\rangle }B^{\langle L\rangle }.
\end{align}
These operators are relevant in the presence of explicit or spontaneous parity violation, such as in the case of magnetically charged objects. Extending this framework to extremal dyonic black holes~\cite{Barbosa:2025smt,Pereniguez:2025jxq} represents a promising direction for future investigation. 
}
These operators describe the cross-induction in the parity-even and parity-odd sectors, respectively. The second operator provides an effective description of both the gravitationally induced magnetic response and the magnetically induced gravitational response. The existence of a single such operator implies that the cross-responses must be identical, which is consistent with the result obtained in the previous subsection. 

Nevertheless, the logarithmic behavior in the worldline description warrants further investigation and a more sophisticated matching with the bulk calculation.

\section{Discussion}

In this work, we have investigated the tidal deformability of extremal charged black holes within the framework of Einstein-Maxwell Effective Field Theory (EFT). By incorporating leading-order four-derivative corrections, we demonstrated that the well-known ``rigidity" of black holes in General Relativity—characterized by vanishing tidal Love numbers—is broken. Our analysis covered both the vector ($\ell=1$) and parity-odd tensor ($\ell \ge 2$) sectors, revealing a rich phenomenology driven by higher-derivative operators.

A central finding of our study is the presence of non-zero black hole tidal responses that exhibit logarithmic running, appearing as $\ln(r/r_h)$ terms in the field profiles.  For the $\ell \ge 2$ modes, this logarithmic behavior indicates that the tidal Love numbers are not constant but are scale-dependent quantities, consistent with the renormalization group (RG) flow expected in an effective field theory description of point particles. This running suggests that the ``Love number" for extremal black holes in EFT should be interpreted as a running coupling defined at a specific scale, rather than a fixed infrared parameter.

\noindent
{\bf Connecting macroscopic deformability to fundamental constraints:} One physically significant result of our analysis is the connection between the sign of the tidal deformation and fundamental constraints on the UV theory. In the dipole ($\ell=1$) sector, we extracted the electric polarizability $\chi_E$ and found that its sign is not arbitrary. Specifically, the condition $\chi_E > 0$ is guaranteed if the EFT coefficients satisfy the bounds imposed by unitarity and the Weak Gravity Conjecture (WGC), namely $\alpha_1 \ge \frac{1}{4}|\alpha_3|$. This establishes a direct link between the consistency of the microscopic theory (absence of superluminality, decay of extremal states) and the macroscopic ``health" of the black hole's tidal response. It supports the intuition that physically consistent black holes should screen, rather than amplify, external electric fields. Moreover, the logarithmic running in the electric dipole sector is governed exclusively by the combination $4\alpha_1 - \alpha_3$, which is subject to constraints from the WGC and unitarity.

\noindent
{\bf Gravito-Electromagnetic mixing and worldline matching:} For higher multipoles ($\ell \ge 2$), our calculation accounted for the inevitable mixing between gravitational and electromagnetic perturbations in the background of a charged black hole. We observed that this mixing manifests symmetrically in the gauge-invariant logarithmic corrections: the gravitationally induced magnetic response is identical to the magnetically induced gravitational response. We argued that this symmetry finds a natural explanation in the worldline EFT, where such cross-responses are generated by single operators of the form $\int d\tau \mathcal{B}_{\langle L \rangle} \mathcal{B}^{\rm EM}_{\langle L \rangle}$. 

However, the appearance of logarithmic terms in the response introduces significant theoretical subtleties and physical consequences into the EFT matching procedure. In standard linearized gravity, the external tidal field (the source) and the black hole's induced multipole moment (the response) are cleanly separated by their distinct radial fall-offs—typically $r^{\ell}$ versus $r^{-(\ell+1)}$. The presence of $\ln(r/r_h)r^{-(\ell+1)}$ terms fundamentally blurs this separation. These logarithmic corrections, which typically signal a running of the Wilson coefficients, complicate the standard definition of Love numbers through asymptotic matching. Theoretically, this issue mirrors the UV divergences encountered in quantum field theory. One has to calculate divergences in the worldline EFT due to the presence of effective operators in the bulk and define the Love numbers at a specific subtraction scale $\mu$ (or radius $R$). Consequently, the precise mapping between the bulk logarithmic terms and the worldline operators remains non-trivial. These subtleties warrant further investigation to establish a rigorous matching scheme that can fully accommodate the scale-dependent nature of tidal deformability.

The physical consequence of this logarithmic running is profound: the tidal deformability of the black hole can no longer be treated as an intrinsic, universal constant of the object's geometry. Instead, the Love number becomes an environment-dependent coupling. In a realistic astrophysical scenario, such as a binary inspiral, this scale $\mu$ would be associated with the orbital separation. Consequently, the black hole's tidal response will dynamically evolve as the binary inspirals, leaving a distinct, scale-dependent imprint on the phase evolution. While our exact coefficients for the logarithmic terms are unambiguously fixed by the bulk EFT, the finite part of the Love number requires fixing this subtraction scale. Establishing a rigorous matching scheme to map these bulk logarithmic terms to worldline operators, and extracting the observable finite-size effects without ambiguity, is a critical next step for black hole EFTs.

While our results focus on the extremal limit, they serve as a crucial stepping stone for understanding the tidal properties of realistic, near-extremal astrophysical black holes. Future work should naturally extend this analysis to non-extremal Reissner-Nordström or Kerr backgrounds to determine if the constraints on electric polarizability and logarithmic running persist away from extremality. Additionally, investigating the interplay between spin, EFT corrections, and tidal deformability remains an important open problem, as the breaking of spherical symmetry is expected to introduce significantly richer mixing patterns.

In this context, extremal Kerr black holes are of particular interest. Recent studies suggest that they serve as a unique arena for amplifying new physics effects~\cite{Horowitz:2023xyl,Chen:2025sim,Cano:2025ejw}. Furthermore, unlike the static case, the tidal deformations of Kerr black holes exhibit dissipative behaviors that are absent for non-rotating objects~\cite{Charalambous:2021mea,Perry:2024vwz}. Extending our analysis to extremal Kerr is therefore essential, both to test the stability of these objects under tidal deformations and to understand how higher-derivative corrections might be macroscopically amplified in the presence of rotation. 

On the observational front, although the effects computed here are suppressed by the EFT scale and about extremal black holes (which are not common astrophysical black holes), it is natural to expect such logarithmic deformations for neutral black holes. A precise characterization of these deformations is vital for next-generation gravitational wave detectors. A detailed understanding of the waveform systematics introduced by these effects will be necessary to distinguish standard GR black holes from exotic compact objects or scenarios arising in modified gravity.

%In conclusion, the non-vanishing Love numbers of extremal black hole are intricate, scale dependent, and deeply constrained by the fundamental principles governing the ultraviolet completion of gravity.

In conclusion, our analysis demonstrates that the inclusion of higher-derivative EFT corrections fundamentally alters the tidal properties of extremal charged black holes. The well-known vanishing of Love numbers in General Relativity is replaced by a rich, non-zero tidal response that is deeply constrained by the fundamental principles of the ultraviolet completion, such as unitarity and the Weak Gravity Conjecture. Notably, we have shown that these tidal responses are not merely shifted by a constant, but exhibit a logarithmic running. The logarithmic terms found in the static response are signal the classical RG running of worldline Love number operators, in which their origin are the effective operators in the gravitational EFT. These results indicate that black hole Love numbers in gravitational EFT should be treated as scale dependent quantity. Moreover, the logarithms expose the essential role of gravito-electromagnetic mixing for strongly charged objects, which motivate a more systematic worldline matching calculation for charged and rotating black holes.

\subsubsection*{Acknowledgments}
We thank Calvin Y.-R. Chen, Valerio De Luca, Lam Hui, Justin Khoury and Luca Santoni for useful and interesting discussions. Special thanks to Valerio De Luca for useful comments on the draft. The work of S.W. is supported by APRC-CityU New Research Initiatives/Infrastructure Support from Central. The work of T.N. is supported by JSPS KAKENHI Grant No. JP22H01220, MEXT
KAKENHI Grant No. JP21H05184 and No. JP23H04007, and JGC-Saneyoshi Scholarship Foundation.

\begin{appendices}

%-------------------------------------------------------------------------------
\section{Electromagnetic polarizability in the background of extremal black holes (ignoring gravitational perturbations)} \label{app:A}
In this section, we study an extremal black hole background as described by the theory in Eq.~\eqref{eqn:action}. We introduce a weak external electromagnetic field and analyze the electromagnetic response, neglecting gravitational back-reaction. This approximation is adopted to facilitate a direct comparison with both the electromagnetic calculation in the previous section and the E\&M EFT analysis in~\cite{Hui:2020xxx}, both of which also omit gravitational perturbations. The gravitational back-reaction effects will be addressed separately in the subsequent section.

We conveniently decompose the electromagnetic gauge field into background and perturbation parts as $ A_\mu = \bar{A}_\mu + a_{\mu}$, where $\bar{A}_\mu$ represents the background gauge field associated with the black hole metric in Eq.~\eqref{eqn:BHmetric}, and $a_{\mu}$ denotes the perturbation around this background. We solve the non-linear E\&M equation from~\eqref{eqn:action}
\begin{align}  \label{eqn:Feom}
    \nabla_{\mu}\left(F^{\mu\nu }- 8 \alpha_1 \kappa^4 F^2F^{\mu\nu } -8\alpha_2 \kappa^4 \tilde{F}F \tilde{F}^{\mu\nu} - 4\alpha_3 \kappa^2 F_{\rho\sigma} W^{\mu\nu \rho\sigma}   \right) =0,
\end{align}
perturbatively in $\alpha_i$ and up to linear order in $a_{\mu}$ since we are treating~\eqref{eqn:action} as an effective field theory and look for linear response.

Due to spherical symmetry, we expand the perturbation in terms of spherical harmonics as
\begin{align}
       a_\mu &= \sum_{\ell, m} \left(\begin{array}{c}
a^{\ell m}_{0}  \\
a^{\ell m}_{r} \\
a^{(L)\ell m} \nabla_A   +a^{(T)\ell m} \epsilon_A^{\;\; B} \nabla_B  
\end{array}\right) Y^m_{\ell}
\end{align}
ensuring that different $(\ell,m)$ modes decouple at linear order. Hereafter, we will omit the explicit indices $(\ell,m)$ unless necessary. Additionally, at linear order, the parity-even sector  $(a_0, a_r, a^{(L)})$ and the parity odd sector $a^{(T)}$ decouple, allowing us to treat these two sectors independently.

\subsection{Parity even sector, electric susceptibility of extremal black hole} \label{sec:evenpureEM}
We first address the parity-even sector $(a_0, a_r, a^{(L)})_{\ell m}$. Under the gauge transformation $A_{\mu} \to A_{\mu} + \partial_{\mu}\Lambda$, these components transform as follows:
\begin{align}
    \delta a_0 = \dot{\Lambda}, \quad \delta a_r =\Lambda' ,\quad  \delta a^{(L)} = \Lambda,
\end{align}
for each $(\ell, m)$. Therefore it is consistent to work with the gauge 
\begin{align}
    a^{(L)} =0. 
\end{align}
To facilitate our analysis, it is convenient to introduce the combination
\begin{align}
    {\cal Q}_{\ell m}=(\dot{a_r}-a_0')_{\ell m},
\end{align}
since $\sum_{\ell m}Q_{\ell m}Y^m_{\ell} = \delta F_{tr}$ precisely captures perturbations in the radial component of the electric field. Solving~\eqref{eqn:Feom} perturbatively in $\{\alpha_i \}$, we systematically expand ${\cal Q}$ as:
\begin{align}
    {\cal Q} = {\cal Q}^{(0)} +{\cal Q}^{(1)} + \dots ,
\end{align}
where ${\cal Q}^{(1)}$ denotes first order terms in $\{\alpha_i \}$. At leading order, ${\cal Q}^{(0)}$ satisfies the equation
\begin{align} \label{eqn:Q0}
    \frac{r^6}{(r-r_h)^2} \ddot{\cal Q}^{(0)} - \partial_r\left(r^2(r-r_h)^2 \partial_r{\cal Q}^{(0)}\right) + \left( (\ell^2 + \ell -2)r^2 + 2r_h^2 \right){\cal Q}^{(0)} =0. 
\end{align}
In the static limit, $\dot{\cal Q}=0$, the general solution to this equation is given by
\begin{align} \label{eqn:evenQgensol}
    c_1 \left(\frac{1}{x} +\frac{1 }{\ell x^2}\right)(x-1)^\ell  + c_2 \frac{(\ell+1)x -1}{(2\ell+1)(\ell+1)x^2}(x-1)^{-\ell -1}.
\end{align}
where $x = \frac{r}{r_h}.$ The requirement of regularity at the horizon $x=1$ imposes $c_2=0$, yielding the zeroth order solution:
\begin{align}
    {\cal Q}^{(0)} =  C_{\cal Q} \left(\frac{1}{x} +\frac{1 }{\ell x^2}\right)(x-1)^\ell,
\end{align}
where $ C_{\cal Q}$ is a constant that quantifies the strength of the external electric field. This solution aligns precisely with the established results describing electric field perturbations in an extremal black hole spacetime.

The first order perturbation ${\cal Q}^{(1)}$, which is linear in the parameters $\{\alpha_i\}$, satisfies a similar differential equation as Eq.~\eqref{eqn:Q0}, now supplemented with a source term determined by the leading-order solution ${\cal Q}^{(0)}$. Imposing regularity at the horizon, the resulting solution exhibits the following behavior:
\begin{align*}
     \ell \ge 2:  \mbox{ there is always a term of the form  } \frac{\ln(r/r_h)}{r^{\ell+2}}.
\end{align*}
For instance, the $\frac{1}{r^{\ell+2}}$ tail in ${\cal Q}$  for $\ell=2,3,4$ are 
\begin{align*}
\ell =2 :\quad  &  C_{\cal Q} \frac{\kappa^2}{r_h^2}3 \frac{  7824 \alpha_1-235 \alpha_3+240 (\alpha_3-20 \alpha_1) \ln x }{25  x^4} , \\
\ell=3 :\quad &   C_{\cal Q} \frac{\kappa^2}{r_h^2} 16 \frac{261040\alpha_1- 12593 \alpha_3 + 1008(7 \alpha_3 -148  \alpha_1) \ln x }{735x^5} ,  \\
\ell =4 :\quad & C_{\cal Q} \frac{\kappa^2}{r_h^2}\frac{23912312 \alpha_1 - 1111131 \alpha_3 + 20160(27\alpha_3 - 584\alpha_1)\ln x}{1134 x^6}   .
\end{align*}
where $x=\frac{r}{r_h}$. The presence of a term proportional to $\frac{\ln(r/r_h)}{r^{\ell+2}}$ obstructs the asymptotic matching with Eq.~\eqref{eqn:radialE}, which is crucial for extracting the Love number.

This observation is consistent with the findings of \cite{Barbosa:2025uau}, which reported the generic appearance of a $\ln\left(\frac{r}{r_h} \right)$ factor accompanying the ${r^{-\ell-2}}$ falloff. There, this behavior is interpreted as a manifestation of the logarithmic running of the Love numbers (although their analysis is based on solving  $\nabla_\mu F^{\mu\nu}=0$ in the higher derivative corrected background of an extremal black hole). In details, a more rigorous matching with the worldline EFT is needed to definitively confirm this interpretation for strongly charged objects. We will demonstrate in the next section that gravitational perturbations cannot be ignored. 

Despite this complication, the $\ell = 1$ sector does not exhibit any logarithmic behavior or the form $\ln\left(\frac{r}{r_h} \right)$ if gravitational perturbations are ignored. For this reason, we will focus our analysis on this sector in what follows. At $\ell=1$,
\begin{align}
    {\cal Q}^{(1)}_{\ell =1} &=  C_{\cal Q} \frac{\kappa^2}{r_h^2} \bigg[-\frac{64}{3} \alpha _1\ln \left( 1 -\frac{1}{x}\right) -\frac{64 \alpha_1 }{3 x} +\frac{   8\alpha_3-80 \alpha _1+ 64 \alpha _1 \ln \left( 1 -\frac{1}{x}\right)}{3 x^2} \nonumber \\
   &\qquad\qquad \quad +\frac{320 \alpha _1+24 \alpha _3  }{5 x^3}-\frac{560 \alpha_1+34 \alpha_3  }{5 x^4}-\frac{16 \alpha_3  }{x^5} + \frac{288 \alpha_1  + 46 \alpha_3} {3 x^6} \bigg] \nonumber \\
   & = C_{\cal Q} \frac{\kappa^2}{r_h^2} \bigg[ \frac{8 \alpha_3 - 48  \alpha_1}{3x^2} + \frac{8(280 \alpha_1 + 27 \alpha_3)}{45x^3}+ {\cal O}(x^{-4}) \bigg] .
\end{align}
Matching to the $\frac{1}{r^{\ell+2}}$ falloff in Eq.~\eqref{eqn:radialE}, we obtain the electric polarizability
\begin{align} \label{eqn:chiE1}
    \chi_E  \approx \frac{4 \kappa^2}{15 r_h^2 }(280\alpha_1+ 27\alpha_3).
\end{align}
To ensure a non-negative polarizability, $\chi_E \ge0$, the EFT coefficients must satisfy
\begin{align} \label{eqn:alphaEpure}
    \alpha_1 \ge -\frac{27}{280}\alpha_3.
\end{align}
Remarkably, this condition is automatically met if the theory satisfies the Weak Gravity Conjecture or the unitarity bound from scattering amplitudes, which require $\alpha_1 \ge \frac{1}{4}|\alpha_3|$. Thus,  {\it the electric polarizability of an extremal black hole in this EFT is guaranteed to be positive } under these consistency conditions (if gravitational perturbations can be ignored for strongly charged objects).

\subsection{Parity odd sector, magnetic susceptibility}
Having examined the parity-even sector, we now turn to the parity-odd sector to complete the analysis. 

The parity odd sector involves only $a^{(T)}$ and it is gauge invariant. We perform the same perturbative (in $\{\alpha_i\}$) analysis as the parity even sector,
\begin{align}
    a^{(T)} = a^{(0)} + a^{(1)}+ \dots ,
\end{align}
where $a^{(1)}$ denotes the first order terms in $\{\alpha_i\}$. The leading order perturbation $a^{(0)}(t,r)$ satisfies   
\begin{align} \label{eqn:oddeq}
     \frac{r^2}{(r^2-r_h^2)} \ddot{a}^{(0)} - \partial_r \left( \frac{(r^2-r_h^2)}{r^2}\partial_r a^{(0)}\right) + \frac{\ell(\ell+1)}{r^2} a^{(0)} =0,
\end{align}
and the most general static solution, $\dot{a}^{(0)}=0$, is given by 
\begin{align} \label{eqn:oddagensol}
   c_1  \left(x+ \frac{ 1}{\ell} \right) (x-1)^\ell +c_2 \left(x- \frac{1}{\ell +1}\right)(x-1)^{-\ell-1},
\end{align}
where $x=\frac{r}{r_h}$.  \footnote{ Note that $x^2{\cal Q}^{(0)}$ in the parity even~\eqref{eqn:evenQgensol} sector and $a^{(0)}$ in the parity odd sector~\eqref{eqn:oddagensol} at the leading order satisfy exactly the same differential equation. This is due to the electromagnetic duality of the Maxwell theory in four dimensions~\cite{Hui:2020xxx}. } Regularity at the horizon sets $c_2=0$ and the corresponding solution 
\begin{align}
   a^{(0)}  =   C_a  \left(x+ \frac{ 1}{\ell} \right) (x-1)^\ell,
\end{align}
represents an external magnetic field with strength quantified by $C_a$. 

The first-order perturbation $a^{(1)}$, linear in the parameters $\{\alpha_i\}$, satisfies a sourced differential equation governed by the same differential operator as in Eq.~\eqref{eqn:oddeq}, with a source term determined by the leading-order solution $a^{(0)}$. After imposing regularity at the horizon, the solutions for $a^{(1)}$ share a similar behavior as the parity even sector: 
\begin{align*}
    \ell \ge 2:  \mbox{ there is always a term }  \frac{\ln(r/r_h)}{r^{\ell}}.
\end{align*} 
For instance 
\begin{align*}
    \ell =2 :\quad  &  C_a \frac{\kappa^2}{r_h^2} 8\frac{ -1554 \alpha_1+918 \alpha_2+305 \alpha_3+ ( 1080\alpha_1-720 \alpha_2-180 \alpha_3) \ln x }{25  x^2}, \\
\ell=3 :\quad &   C_a \frac{\kappa^2}{r_h^2} 64 \frac{ -40564 \alpha_1+23504 \alpha_2+7203 \alpha_3+ (23184\alpha_1 -13440 \alpha_2 -4116 \alpha_3) \ln x }{735x^3}, \\
\ell =4 :\quad & C_a \frac{\kappa^2}{r_h^2} 4 \frac{-1885657\alpha_1+1004354 \alpha_2+345870 \alpha_3 + (927360\alpha_1 - 493920\alpha_2 - 170100\alpha_3)\ln x}{567 x^4}  .
\end{align*}
As in the parity-even sector, the appearance of a $\frac{\ln(r/r_h)}{r^{\ell}}$ term precisely at the $r^{-\ell}$ order obstructs the extraction of the Love number through asymptotic matching. Also, these are usually interpreted as the running of the magnetic susceptibilities. 

Therefore, we only focus on the $\ell=1$ sector where logarithmic runnings are absent, 
\begin{align}
     a^{(1)}_{\ell=1}& = C_a \frac{\kappa^2}{r_h^2} \bigg[ \frac{64}{3} \left(\alpha _1-2 \alpha _2\right) x^2 \ln \left( 1 -\frac{1}{x}\right) +\frac{64}{3} \left(\alpha _1-2 \alpha _2\right) x  \nonumber \\
     &\qquad \qquad  - \frac{8}{3}  \left(2 \alpha _1+8 \alpha _2-\alpha _3+8 \left(\alpha _1-2 \alpha _2\right)  \ln \left( 1 -\frac{1}{x}\right)\right) \nonumber \\
    &\qquad \qquad  -\frac{8 \left(24 \alpha _1-32 \alpha _2-7 \alpha _3\right)}{5 x} +\frac{8 \left(42 \alpha _1+24 \alpha _2-26 \alpha _3\right)}{15 x^2}
     \bigg] \nonumber \\
     & =  C_a \frac{\kappa^2}{r_h^2} \bigg[ \frac{8}{3} \left(\alpha _3-6 \alpha _1\right)-\frac{8 \left(136 \alpha _1-128 \alpha _2-63 \alpha _3\right)}{45 x} + {\cal O}(x^{-2})\bigg].
\end{align}
With this solution, one can match with Eq.~\eqref{eqn:FABodd} and~\eqref{eqn:FABa} to figure out 
\begin{align} \label{eqn:chiM1}
    \chi_M \approx \frac{4 \kappa^2}{15 r_h^2}\left(-136\alpha_1 + 128 \alpha_2 + 63\alpha_3 \right).
\end{align}
From the result of $\chi_E$ in the parity even sector~\eqref{eqn:chiE1}, we have 
\begin{align}
  \chi_M+\chi_E \approx  \frac{8}{15} (72 \alpha_1+64 \alpha_2+45 \alpha_3).
\end{align}
This quantity is physically meaningful since the causality constraint in Eq.~\eqref{eqn:causalconstr} requires $\chi_E + \chi_M \ge 0$. Since we have $\alpha_2 \ge0$ from the WGC~\eqref{eqn:WGCconstraints}, the causality constraint would be sufficiently satisfied if $72 \alpha_1+45 \alpha_3 \ge 0$. It can be compared with the WGC constraint, $4\alpha_1 -|\alpha_3| \ge 0$. 
\begin{figure}[h]
    \centering
    \includegraphics[width=0.35\linewidth]{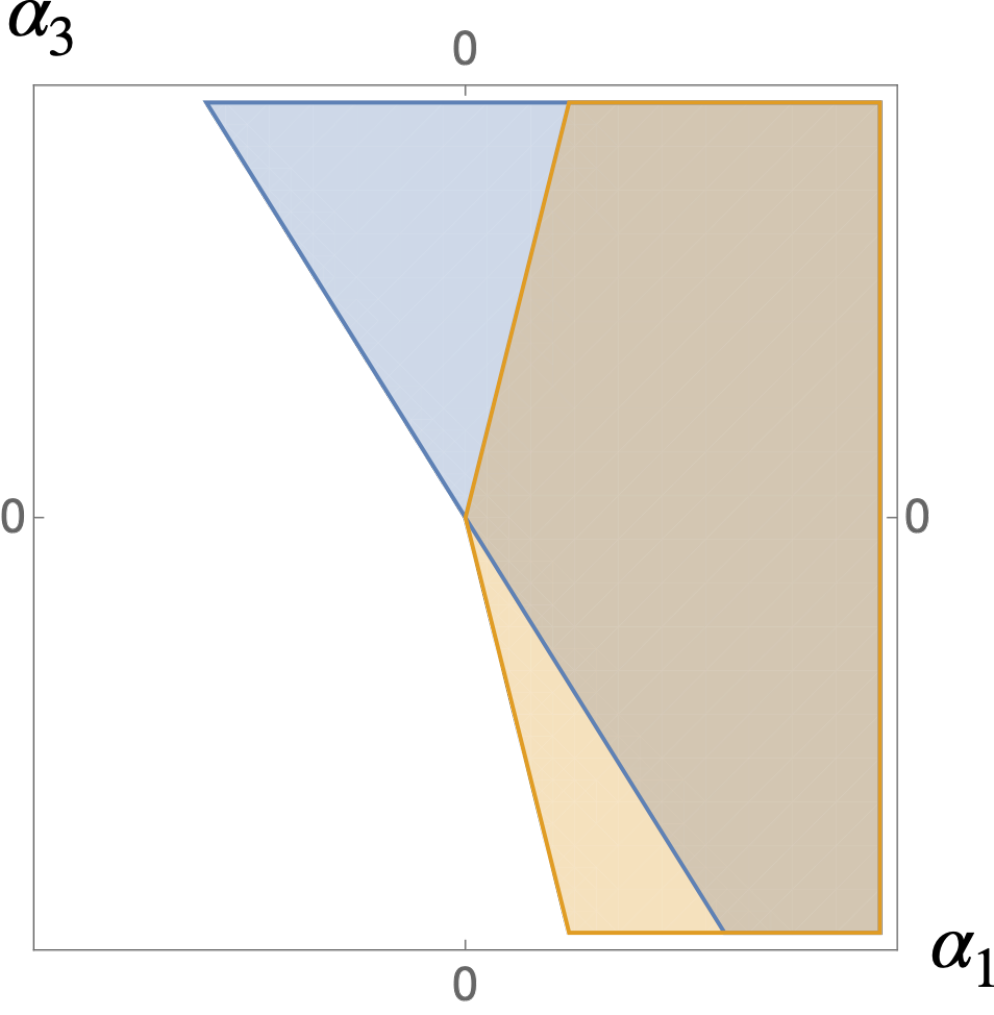}
    \caption{Blue region: the constraint from $72 \alpha_1+45 \alpha_3 \ge 0$. Brown region: $4\alpha_1 -|\alpha_3| \ge 0$.}
    \label{fig:chiEChiM}
\end{figure}

As shown in figure~\eqref{fig:chiEChiM}, it may seem that the constraint imposed by the WGC does not necessarily imply the constraint required by causality, which arises from the condition of positive electromagnetic polarizability. However, we should emphasize that this result cannot be taken seriously since we only considered small E\&M perturbations in the background of corrected extremal black hole. When we compare this result with the main text, the mixing of gravitational and electromagnetic perturbations leads to non-negligible contributions. This comparison also demonstrates the subtleties that arises in the process of world-line EFT matching for highly charged objects. 
\end{appendices}

\bibliographystyle{utphys}
\bibliography{WGCTLN.bib}

\end{document}